\documentclass[11pt,british,english]{article}
\pdfoutput=1
\usepackage[T1]{fontenc}
\usepackage[latin9]{inputenc}
\usepackage{geometry}
\usepackage{color}
\usepackage{amsmath}
\usepackage{amssymb}
\usepackage{subfig}
\usepackage{esint}
\usepackage{cite}
\usepackage{float}
\usepackage[colorlinks=true,linkcolor=blue, citecolor=red, urlcolor=blue, bookmarks]{hyperref}

\makeatletter

\author{Dávid X. Horváth$^{1}$, Luca Capizzi$^{1}$, and Pasquale Calabrese$^{1,2}$~\\
 $^{1}${\small{}{}SISSA and INFN Sezione di Trieste, via Bonomea
265, 34136 Trieste, Italy}.\\
 $^{2}${\small{}{}International Centre for Theoretical Physics (ICTP),
Strada Costiera 11, 34151 Trieste, Italy.} }

\usepackage{babel}

\usepackage{babel}

\makeatother

\begin{document}

\title{\textbf{U(1) symmetry resolved entanglement in free 1+1 dimensional
field theories via form factor bootstrap}}
\maketitle
\begin{abstract}
We generalise the form factor bootstrap approach to integrable field
theories with $U(1)$ symmetry to derive matrix elements of composite
branch-point twist fields associated with symmetry resolved entanglement
entropies. The bootstrap equations are solved for the free massive
Dirac and complex boson theories, which are the simplest theories
with $U(1)$ symmetry. We present the exact and complete solution
for the bootstrap, including vacuum expectation values and form factors
involving any type and arbitrarily number of particles. The non-trivial
solutions are carefully cross-checked by performing various limits
and by the application of the $\Delta$-theorem. An alternative and
compact determination of the novel form factors is also presented.
Based on the form factors of the $U(1)$ composite branch-point twist
fields, we re-derive earlier results showing entanglement
equipartition for an interval in the ground state of the two models. 
\end{abstract}

\baselineskip 18pt \thispagestyle{empty} \newpage{}

\tableofcontents{}

\section{Introduction}

Symmetries play an inevitable role in modern physics and deeply interlace
the whole discipline in versatile ways. Restricting to merely a practical
point of view,  it is generally a very fruitful idea to exploit the additional structures imposed by symmetry for almost any type of physical object. 
Interestingly, such an idea has only recently been investigated in the study of entanglement of extended quantum systems. 
As widely known, when a system is in a pure state, the bipartite entanglement
of a subsystem $A$ may be quantified by the R\'enyi entanglement entropies \cite{vNE1,vNE2,vNE3,vNE4}
\begin{equation}
S_{n}=\frac{1}{1-n}\ln\text{Tr}\rho_{A}^{n}\,,
\end{equation}
defined in terms of the reduced density matrix (RDM) $\rho_{A}$ of the subsystem $A$.
From those, in the replica limit $n\rightarrow1$ the von Neumann entropy 
\begin{equation}
S=-\text{Tr}\rho_{A}\ln\rho_{A}.
\end{equation}
is obtained. 
However, the knowledge of R\'enyi entropies for different $n$ contains more information than the simple $S$, as, e.g., it provides the entire spectrum of the reduced 
density matrix $\rho_A$ \cite{cl-08,ace-18}.

The recent and explicit idea of considering generally the internal
structure of entanglement associated with symmetry was put forward
in Refs. \cite{lr-14,Greiner,gs-18,SRENegativity}. 
In a symmetric state, the  conserved charge corresponding to the symmetry $\hat{Q}$ commutes with density matrix; 
under general circumstances also the restriction of $\hat{Q}$ to the subsystem $\hat{Q}_{A}$, commutes with the RDM as 
\begin{equation}
[\rho_{A},\hat{Q}_{A}]=0\,.\label{eq:Commutation}
\end{equation}
Such commutation implies that $\rho_{A}$ is block-diagonal, each block corresponding to an eigenvalue of $\hat{Q}_{A}$. 
This fact has the important consequence that the
Rényi and von Neumann entropies can be decomposed according to the
symmetry sectors of $\hat{Q}_{A}$. 
The symmetry resolved R\'enyi and von Neumann entropies are defined as
\begin{equation}
S_{n}(q_{A})=\frac{1}{1-n}\ln\left[\frac{\mathcal{Z}_{n}(q_{A})}{\mathcal{Z}_{1}^{n}(q_{A})}\right]\,,\qquad{\rm and}\qquad S(q_{A})=-\frac{\partial}{\partial n}\left[\frac{\mathcal{Z}_{n}(q_{A})}{\mathcal{Z}_{1}^{n}(q_{A})}\right]_{n=1}\,,\label{eq:vNE}
\end{equation}
in terms of  the symmetry resolved partition sums
\begin{equation}
\mathcal{Z}_{n}(q_{A})=\text{Tr}\left(\rho_{A}^{n}\mathcal{P}(q_{A})\right)\,,
\end{equation}
where $\mathcal{P}(q_{A})$ is the projector onto the sector corresponding to the eigenvalue $q_{A}$.

The calculation of all these symmetry resolved quantities requires in general the diagonalisation of $\rho_A$ and the resolution of the spectrum in the conserved charge. 
An ingenious way to circumvent this difficult path passes through  the charged moments \cite{gs-18}
\begin{equation}
Z_{n}(\alpha)=\text{Tr}\left(\rho_{A}^{n}e^{i\alpha\hat{Q}_{A}}\right),\label{eq:Zn}
\end{equation}
that are nothing but the Fourier transform in of the desired partition sums, i.e. \cite{gs-18}
\begin{equation}
\mathcal{Z}_{n}(q_{A})=\text{Tr}\left(\rho_{A}^{n}\mathcal{P}(q_{A})\right)=\int_{-\pi}^{\pi}\frac{\mathrm{d}\alpha}{2\text{\ensuremath{\pi}}}Z_{n}(\text{\ensuremath{\alpha}})e^{-i\alpha q_{A}}.\label{eq:CalU1}
\end{equation}
This way of proceeding is particularly powerful for field theoretical calculations in the path-integral formalism. 
In a replica approach, that is reviewed in the following, $\text{Tr}\rho_{A}^{n}$ is a partition function
on an $n$-sheeted Riemann surface $\mathcal{R}_{n}$, obtained by joining cyclically the $n$ sheets along the subsystem $A$ \cite{Replica,cc-04,cc-09}.
In this language, the charged moments (\ref{eq:Zn}) correspond to introducing an Aharonov-Bohm
flux on one of the sheets of $\mathcal{R}_{n}$ \cite{gs-18}. 
%
Symmetry resolved entropies have been studied in various field theoretical
contexts such as conformal field theories (CFTs) \cite{gs-18,lr-14,Equipartitioning,SREQuench,crc-20,eimd-20,bc-20,mbc-21},
free \cite{mdc-20b} and interacting integrable QFTs \cite{Z2IsingShg}, and holographic settings \cite{znm}.
Many results are also known for microscopical models on the lattice \cite{lr-14,Equipartitioning,SREQuench,brc-19,fg-20,FreeF1,FreeF2,mdc-20,ccdm-20,pbc-20,bc-20,HigherDimFermions,mrc-20},
disordered systems \cite{MBL, MBL2,trac-20}, and for non-trivial topological
phases \cite{Topology,Anyons}. Symmetry resolved quantities can also
be measured experimentally \cite{vecd-20}. 
It is worth saying that the charged moments \eqref{eq:Zn} have also been the subject of field theoretical studies  
before their application to symmetry resolved entropies was recognised \cite{bym-13,cms-13,cnn-16,d-16,d-16b,ssr-17,srrc-19}.

A very standard and powerful trick, used a lot in quantum field theory (QFT) context, is to replace the partition function on the $n$-sheeted surfaces
with an $n$-copy version of the original model with specific boundary conditions among the replicated fields \cite{CFH}. 
Crucially, in 1+1 dimension, these boundary conditions can be implemented via 
local fields in the $n$-copy theory that are known as branch-point twist fields \cite{cc-04,ccd-08}.
The moments of $\rho_A$ are eventually equivalent to an appropriate multi-point function of the branch-point twist fields
in the $n$-copy theory. In 2D CFT,  the scaling dimensions of these fields are exactly
known \cite{cc-04,k-87,dixon} and provide their two-point function, i.e. the entanglement entropy of a single interval
for a generic CFT \cite{cc-04}. 
The behaviour of multi-point correlation functions  of the branch-point twist fields are
known for special CFTs \cite{fps-08,cct-09,h-10,cct-11,atc-11,rg-12,rg-12b,dei-18,ctt-14}.

In 2D off-critical integrable and free theories, the form factor (FF) bootstrap allows for the calculation of the matrix elements of
the branch-point twist field \cite{ccd-08,cd-08,cd-09}. 
Although all matrix elements are in principle computable, the multi-point correlation functions at large distances are generically dominated 
by the first few sets of form factors. 
This property has been exploited in integrable QFTs (IQFTs) to obtain predictions for the entanglement entropy in many different physical circumstances 
\cite{cd-09b,cl-11,cdl-12,leviFFandVEV,lcd-13,bcd-15,bcd-15b,bcd-16,bc-16,c-17,cdds-18,cdds-18b,cdds-18c,cdds-18d,clsv-19,clsv-20}.

In 2D CFT,  the symmetry resolved entropies are obtained as multipoint correlations of 
composite branch-point twist fields which fuse the action of the replicas
and of the flux of charge \cite{gs-18}.  These composite twist fields have been recently identified also in some massive theories:
free massive Dirac and complex boson QFT \cite{mdc-20b},  the off-critical Ising and sinh-Gordon theories \cite{Z2IsingShg}.
As shown in our previous work \cite{Z2IsingShg}, form factors of
the composite twist fields can be determined with the bootstrap program,
similarly to the usual branch-point twist fields \cite{ccd-08,cd-08,cd-09}.
Using these form factors, one can obtain a systematic expansion for the correlation functions
of the composite twist fields, which eventually relate to the symmetry resolved entropies.

Our previous work \cite{Z2IsingShg} focused on the symmetry resolution
of the 
discrete $\mathbb{Z}_{2}$ symmetry in the Ising and sinh-Gordon models
\cite{Z2IsingShg}. Here, instead, we initiate the study of continuous
symmetries: we use appropriate bootstrap equations for the composite
branch-point twist fields related to the $U(1)$ symmetry and solve
these equations. For simplicity, we consider the free massive Dirac
theory and the free massive complex boson theory. Although the symmetry
resolution of these theories has already been performed \cite{mdc-20b},
it is extremely instructive to study them from the point of view of
form factor bootstrap, being the simplest examples of theories with
more than one particle species. Indeed, multi-particle theories usually
introduce quite a few technicalities, which can be kept at a minimum
level in free theories. Nevertheless, as we show here, the form factors
of the composite twist fields are non-trivial objects even in free
theories. Hence, our main result is the determination and solution
of the bootstrap equations for these form factors. We then use them
to infer the behaviour of the charged moments when the subsystem is
a long interval. Finally, for completeness, we also compute the symmetry
resolved partition functions and entropies within the two-particle
approximation. 

The paper is structured as follows. In section \ref{sec:ConventionalTwistFields}
the FF approach for conventional branch-point twist fields is briefly
reviewed, and the FF solutions in the free massive Dirac and complex
boson theory are presented. Sections \ref{sec:U1DiracTwistFields}
and \ref{secBoson} are devoted to the FFs of the composite branch-point
twist fields in the Dirac and complex boson theory, respectively.
Starting from the appropriate bootstrap equations the 2-particle FFs
are determined first, from which FFs involving any number of particles
are obtained as well. Our solutions are carefully cross-checked by
studying various limits of the FFs and by applying the $\Delta$-theorem
\cite{delta_theorem}. In section \ref{secAlternative} a compact,
alternative derivation of the 2-particle FFs are presented, which
is based on an explicit diagonalisation in the replica space. Section
\ref{sec:gen} reports general and specific results for $U(1)$ charged
moments that can be deduced from the IQFT structure. 
The leading and sub-leading contributions of the symmetry resolved
entanglement are explicitly calculated in section \ref{sec:CalculatingEntropies}.
We finally conclude in section \ref{sec:Conclusions}, which is followed
by the appendices containing some details on the analytic continuation
of the replica index $n\rightarrow1$ (appendix \ref{sec:AppendixA-AnalContForfDB}),
and the determination of the vacuum expectation value (VEV) of the
composite branch-point twist field (appendix \ref{sec:AppendixB-VEVDirac}
and \ref{sec:AppendixC-VEVBoson}) .

\section{Form factors of the branch-point twist fields in integrable models\label{sec:ConventionalTwistFields}}

A primary focus of this work is on the composite branch-point
twist field and their FFs in free models, nevertheless, before discussing
these objects, it is natural to review the standard branch-point twist
fields and the corresponding FFs first. These FFs are an important
reference point for our later investigations. Moreover this brief
overview allows us to conveniently introduce some basic ingredients
of IQFTs, which are essential building blocks in our derivations.
Such central objects in our investigations are the FF bootstrap equations
and the FFs themselves. Following closely the logic of Ref. \cite{ccd-08},
we introduce the bootstrap equations for
branch-point twist field and comment on their solution. Since it takes
little effort and helps keep connection with interacting IQFTs, we
keep the following discussion more general, which thus describes the
case of interacting theories with diagonal but non-trivial scattering
as well.

First of all, let us recall that form factors (FF)
are matrix elements of (semi-)local operators $O(x,t)$ between the
vacuum and asymptotic states, i.e., 
\begin{equation}
F_{\beta_{1},\ldots,\beta_{n}}^{O}(\vartheta_{1},\ldots,\vartheta_{n})=\langle0|O(0,0)|\vartheta_{1},\ldots\vartheta_{n}\rangle_{\beta_{1},\ldots,\beta_{n}}.\label{eq:FF}
\end{equation}
In massive field theories, the asymptotic states are
spanned by multi-particle excitations whose dispersion
relation can be parametrised as $(E,p)=(m_{\beta_{i}}\cosh\vartheta,m_{\beta_{i}}\sinh\vartheta)$,
where $\beta_{i}$ indicates the particle species and
$\vartheta$ is the rapidity of the particle. In such models, any
multi-particle state can be constructed from vacuum state $|0\rangle$
as 
\begin{equation}
|\vartheta_{1},\vartheta_{2},...,\vartheta_{n}\rangle_{\beta_{1},\ldots,\beta_{n}}=A_{\beta_{1}}^{\dagger}(\vartheta_{1})A_{\beta_{2}}^{\dagger}(\vartheta_{2})\ldots.A_{\beta_{n}}^{\dagger}(\vartheta_{n})|0\rangle\:,\label{eq:basis}
\end{equation}
where $A^{\dagger}$s are particle creation operators;
in particular the operator $A_{\beta_{i}}^{\dagger}(\vartheta_{i})$
creates a particle of species $\beta_{i}$ with rapidity $\vartheta_{i}$.
In an IQFT with factorised scattering, the creation and
annihilation operators $A_{\beta_{i}}^{\dagger}(\vartheta)$ and $A_{\beta_{i}}(\vartheta)$
satisfy the Zamolodchikov-Faddeev (ZF) algebra, which, for diagonal scattering, reads 
\begin{eqnarray}
A_{\beta_{i}}^{\dagger}(\vartheta_{i})A_{\beta_{j}}^{\dagger}(\vartheta_{j}) & = & S_{\beta_{i},\beta_{j}}(\vartheta_{i}-\vartheta_{j})A_{\beta_{j}}^{\dagger}(\vartheta_{j})A_{\beta_{i}}^{\dagger}(\vartheta_{i})\:,\nonumber \\
A_{\beta_{i}}(\vartheta_{i})A_{\beta_{j}}(\vartheta_{j}) & = & S_{\beta_{i},\beta_{j}}(\vartheta_{i}-\vartheta_{j})A_{\beta_{j}}(\vartheta_{j})A_{\beta_{i}}(\vartheta_{i})\:,\nonumber \\
A_{\beta_{i}}(\vartheta_{i})A_{\beta_{j}}^{\dagger}(\vartheta_{j}) & = & S_{\beta_{i},\beta_{j}}(\vartheta_{j}-\vartheta_{i})A_{\beta_{j}}^{\dagger}(\vartheta_{j})A_{\beta_{i}}(\vartheta_{i})+\delta_{\beta_{i},\beta_{j}}2\pi\delta(\vartheta_{i}-\vartheta_{j}),\label{eq:ZF}
\end{eqnarray}
where $S_{\beta_{i},\beta_{j}}(\vartheta_{i}-\vartheta_{j})$ denotes
the two-particle S-matrix of the theory that describes the scattering of particles $\beta_{i}$ and $\beta_{j}$. 

Turning to an $n$-copy IQFT one describes the scattering
between particles in different and in the same copies as
\begin{equation}
\begin{split}S_{(\beta_{i}\mu_{i}),(\beta_{j}\mu_{j})}(\vartheta) & =1,\;\qquad i,j=1,...,n\text{ and }\mu_{i}\neq\mu_{j},\\
\quad\quad\quad S_{(\beta_{i}\mu_{i}),(\beta_{j}\mu_{j})}(\vartheta) & =S_{\beta_{i},\beta_{j}}(\vartheta),\;\;\;i,j=1,..,n\text{ and }\mu_{i}=\mu_{j},
\end{split}
\end{equation}
where we introduced the replica index $\mu_{i}$ , which takes values
from $1$ to $n$.  As known, in the $n$-copy QFT the
branch-point twist fields are related to the symmetry $\sigma\Psi_{i}=\Psi_{i+1}$,
with $n+i\equiv i$. When a twist field $\mathcal{T}$ (or ${\cal T}_{n}$)
is inserted in a correlation function, its action can be summarised
as 
\begin{equation}
\begin{split}\Psi_{i}(y)\mathcal{T}(x)=\mathcal{T}(x)\Psi_{i+1}(y) & \qquad x<y,\\
\Psi_{i}(y)\mathcal{T}(x)=\mathcal{T}(x)\Psi_{i}(y) & \qquad x>y.
\end{split}
\end{equation}
In a similar way, one can also define $\tilde{\mathcal{T}}$ with action 
\begin{equation}
\begin{split}\Psi_{i}(y)\mathcal{\tilde{\mathcal{T}}}(x)=\mathcal{\tilde{\mathcal{T}}}(x)\Psi_{i-1}(y) & \qquad x{>y},\\
\Psi_{i}(y)\mathcal{\tilde{\mathcal{T}}}(x)=\tilde{\mathcal{T}}(x)\Psi_{i}(y) & \qquad x<y.
\end{split}
\end{equation}
The bootstrap equations for branch-point twist fields
are natural modifications of the form factor bootstrap equations for
conventional fields \cite{bergkarowski,kirillovsmirnov,FFAxioms}.
Exploiting the relation of the twist fields to the symmetry $\sigma\Psi_{i}=\Psi_{i+1}$,
the corresponding bootstrap equations can be written as \cite{ccd-08}
\begin{eqnarray}
 &  & F_{k}^{\mathcal{T}|\underline{(\beta\mu)}}(\underline{\vartheta})=S_{(\beta_{i}\mu_{i}),(\beta_{i+1}\mu_{i+1})}(\vartheta_{i,i+1})F_{k}^{\mathcal{T}|...(\beta_{i+1}\mu_{i+1}),(\beta_{i}\mu_{i})...}(\ldots\vartheta_{i+1},\vartheta_{i},\ldots),\label{eq:FFAxiom1}\\
 &  & F_{k}^{\mathcal{T}|\underline{(\beta\mu)}}(\vartheta_{1}+2\pi i,\vartheta_{2},\ldots,\vartheta_{k})=F_{k}^{\mathcal{T}|(\beta_{2}\mu_{2}),...,(\beta_{k}\mu_{k}),(\beta_{1}\hat{\mu}_{1})}(\vartheta_{2},\ldots,\vartheta_{n},\vartheta_{1}),\label{eq:FFAxiom2}\\
 &  & -i\underset{\vartheta_{0}'=\vartheta_{0}+i\pi}{{\rm Res}}F_{k+2}^{\mathcal{T}|(\beta \mu),(\bar{\beta} \mu),\underline{(\beta\mu)}}(\vartheta_{0}',\vartheta_{0},\underline{\vartheta})=F_{k}^{\mathcal{T}|\underline{(\beta\mu)}}(\underline{\vartheta}),\label{eq:FFAxiom3}\\
 &  & -i\underset{\vartheta_{0}'=\vartheta_{0}+i\pi}{{\rm Res}}F_{k+2}^{\mathcal{T}|(\beta \mu),(\bar{\beta} \hat{\mu}),\underline{(\beta\mu)}}(\vartheta_{0}',\vartheta_{0},\underline{\vartheta})=-\prod_{i=1}^{k}S_{(\beta \hat{\mu}),(\beta_{i}\mu_{i})}(\vartheta_{0i})F_{k}^{\mathcal{T}|\underline{(\beta\mu)}}(\underline{\vartheta}),\nonumber \\
 &  & -i\underset{\vartheta_{0}'=\vartheta_{0}+i\bar{u}_{\beta\gamma}^{\delta}}{{\rm Res}}F_{k+2}^{\mathcal{T}|(\beta \mu),(\gamma \mu'),\underline{(\beta\mu)}}(\vartheta_{0}',\vartheta_{0},\underline{\vartheta})=\delta_{\mu\mu'}\Gamma_{\beta\gamma}^{\delta}F_{k+1}^{\mathcal{T}|(\delta\mu),\underline{(\beta\mu)}}(\vartheta_{0},\underline{\vartheta}),\label{eq:FFAxiom4}
\end{eqnarray}
where $\vartheta_{ij}=\vartheta_{i}-\vartheta_{j}$, $\underline{\vartheta}$
and $\underline{(\beta\mu)}$ are shorthands for $\vartheta_{1},\vartheta_{2},...,\vartheta_{k}$
and $(\beta_{1}\mu_{1}),(\beta_{2}\mu_{2})....,(\beta_{k}\mu_{k})$
respectively; $\hat{\mu}=\mu+1$ and $\bar{\beta}$ denotes the anti-particle
of species $\beta$. In the last equation $\bar{u}_{\beta\gamma}^{\delta}$
refers to the position of the pole of the bound state of the particle
$\delta$ in the S-matrix $S_{\beta,\gamma}$ and $i\left(\Gamma_{\beta\gamma}^{\delta}\right)^{2}$
is the corresponding pole-strength. Furthermore, relativistic invariance
implies 
\begin{equation}
F_{k}^{\mathcal{T}|\underline{(\beta\mu)}}(\vartheta_{1}+\Lambda,\ldots,\vartheta_{k}+\Lambda)=e^{s\Lambda}F_{k}^{\mathcal{T}|\underline{(\beta\mu)}}(\vartheta_{1},\ldots,\vartheta_{k}),\label{eq:RelInv}
\end{equation}
where $s$ is the Lorentz spin of the operator, which is zero for
the branch-point twist fields. The free Dirac and massive complex
boson theories have no bound states; consequently Eqs. (\ref{eq:FFAxiom1})-(\ref{eq:FFAxiom3})
and (\ref{eq:RelInv}) give all the constraints for form factors of
the branch-point twist fields.

The above bootstrap equations have many inequivalent solutions. Some additional information
is generally needed to sort out a given physical solution.
In this respect a very useful tool is the $\Delta$-theorem \cite{delta_theorem}. 
Such a theorem states that when at some length scale $R$ the theory is described by a CFT,
the difference between the conformal weight of an operator $O$ and its conformal weight in the infrared (IR) limit is
\begin{equation}
D(R)-\Delta^{IR}=-\frac{1}{4\pi\langle O\rangle}\int_{x^{2}>R}\mathrm{d^{2}}x\langle\Theta(x)O(0)\rangle_{c},\label{eq:DeltaTheoreM1}
\end{equation}
where $\Theta$ is the trace of the stress-energy tensor. 
Using the form factors to write the correlation function in the integral above, we have
\begin{equation}
D(r)-\Delta^{IR}=-\frac{1}{2\left\langle O\right\rangle }\sum_{n=1}^{\infty}\sum_{\beta_{1},...,\beta_{n}}\int\frac{\mathrm{d}\vartheta_{1}...\mathrm{d}\vartheta_{n}}{(2\pi)^{n}n!}\frac{e^{-rE_{n}}(1+E_{n}r)}{m^{2}E_{n}^{2}}F_{n}^{\Theta|\underline{\beta}}\left(\vartheta_{1},\dots,\vartheta_{n}\right)F_{n}^{O|\underline{\beta}}\left(\vartheta_{1},\dots,\vartheta_{n}\right)^{*}\,,\label{eq:DeltaTheoreM2}
\end{equation}
where $m$ is a mass scale $r=Rm$ and $mE_{n}$ are the $n$-particle
energies with $E_{n}=\sum_{k=1}^{n}m_{\beta_{k}}\cosh\vartheta_{k}/m$.
For massive theories, the conformal weights in the IR are zero. Hence taking $r=0$ in (\ref{eq:DeltaTheoreM2}) gives
the conformal dimension of the operator $O$ as
\begin{equation}
\Delta^{UV}=-\frac{1}{2\left\langle O\right\rangle }\sum_{n=1}^{\infty}\sum_{\beta_{1},...,\beta_{n}}\int\frac{\mathrm{d}\vartheta_{1}...\mathrm{d}\vartheta_{n}}{(2\pi)^{n}n!}m^{-2}E_{n}^{-2}F_{n}^{\Theta|\underline{\beta}}\left(\vartheta_{1},\dots,\vartheta_{n}\right)F_{n}^{O|\underline{\beta}}\left(\vartheta_{1},\dots,\vartheta_{n}\right)^{*}\,.\label{eq:DeltaTheoreM3}
\end{equation}
Consequently, given a set of FFs satisfying the axioms \eqref{eq:FFAxiom1}-\eqref{eq:FFAxiom4}
and \eqref{eq:RelInv}, the $\Delta$-theorem allows us to identify to which physical operator it corresponds (as long as its conformal dimension is known). 
Although in principle the $\Delta$-theorem sum rule requires the knowledge of infinitely many terms and FFs, it is usually
a very rapidly converging sum; therefore the first few-particle FFs are enough to identify the UV conformal dimension. 
Finally, we notice that the UV dimension imposes also a constraint on the growth of FFs for large rapidities. 
When one rapidity of the FF goes to $\pm\infty$, its growth is utmost $e^{y_{O}\left|\vartheta_{i}\right|}$
with $y_{o}\leq\Delta_{O}$, where $O$ is the operator of interest.

The FFs of various fields are usually found first solving Eqs. \eqref{eq:FFAxiom1}-\eqref{eq:FFAxiom4}
and then removing the possible ambiguities using the $\Delta$-theorem
sum rule. The starting point is the minimal form factor $F_{\text{min}}^{\mathcal{T}|(\beta j),(\gamma k)}(\vartheta,n)$
defined as the solution of the first two equations, Eqs. (\ref{eq:FFAxiom1})
and (\ref{eq:FFAxiom2}), i.e., it satisfies 
\begin{equation}
F_{\text{min}}^{\mathcal{T}|(\beta k),(\gamma j)}(\vartheta,n)=F_{\text{min}}^{\mathcal{T}|(\gamma j),(\beta k)}(-\vartheta,n)S_{(\beta k),(\gamma j)}(\vartheta)=F_{\text{min}}^{\mathcal{T}|(\gamma j),(\beta k+1)}(2\pi i-\vartheta,n)\,.
\end{equation}
As a consequence we have
\begin{equation}
\ \begin{split}F_{\text{min}}^{\mathcal{T}|(\beta i),(\gamma i+k)}(\vartheta,n)= & F_{\text{min}}^{\mathcal{T}|(\beta j),(\gamma j+k)}(\vartheta,n)\quad\forall i,j,k\\
F_{\text{min}}^{\mathcal{T}|(\beta1),(\gamma j)}= & F_{\text{min}}^{\mathcal{T}|(\gamma1),(\beta1)}(2\pi i(j-1)-\vartheta,n)\quad\forall j\neq1\,,
\end{split}
\label{eq:FMinAxioms}
\end{equation}
and hence
\begin{equation}
F_{\text{min}}^{\mathcal{T}|(\beta j),(\gamma k)}=\begin{cases}
F_{\text{min}}^{\mathcal{T}|(\gamma1),(\beta1)}(2\pi i(k-j)-\vartheta,n) & \text{ if }k>j,\\
F_{\text{min}}^{\mathcal{T}|(\beta1),(\gamma1)}(2\pi i(j-k)+\vartheta,n) & \text{otherwise},
\end{cases}
\end{equation}
so that the only independent quantity is $F_{\text{min}}^{\mathcal{T}|(\beta1),(\gamma1)}$.
The latter is determined by  Eq. (\ref{eq:FMinAxioms}) rewritten as 
\begin{equation}
F_{\text{min}}^{\mathcal{T}|(\beta1),(\gamma1)}=F_{\text{min}}^{\mathcal{T}|(\gamma1),(\beta1)}(-\vartheta,n)S_{\beta,\gamma}(\vartheta)=F_{\text{min}}^{\mathcal{T}|(\gamma1),(\beta1)}(-\vartheta+2\pi in,n)\,.
\end{equation}
To solve the last equation, we introduce  $f_{11}(\vartheta)$ satisfying 
\begin{equation}
f_{(\beta1),(\gamma1)}(\vartheta)=f_{(\gamma1),(\beta1)}(-\vartheta)S_{\beta,\gamma}(n\vartheta)=f_{(\gamma1),(\beta1)}(-\vartheta+2\pi i)\,,\label{f11axiom}
\end{equation}
so that
\begin{equation}
F_{\text{min}}^{\mathcal{T}|(\beta1),(\gamma1)}(\vartheta,n)=f_{(\beta1),(\gamma1)}(\vartheta/n)\,.
\end{equation}
Eq. (\ref{f11axiom}) is just the standard equation for minimal form factors
of conventional local operators, but with an S-matrix $S(n\vartheta)$
rather then $S(\vartheta)$. 
Exploiting now the standard parametrisation of $S(\vartheta)$ 
\begin{equation}
S_{\beta,\gamma}(\vartheta)=\exp\left[\int_{0}^{\infty}\frac{\mathrm{d}t}{t}g_{\beta,\gamma}(t)\sinh\frac{t\vartheta}{i\pi}\right],\label{Svsg}
\end{equation}
with some function $g(t)$, the minimal FF is 
\begin{equation}
f_{(\beta1),(\gamma1)}(\vartheta)=\mathcal{N}\exp\left[\int_{0}^{\infty}\frac{\mathrm{d}t}{t}\frac{g_{\beta,\gamma}(t)}{\sinh nt}\sin^{2}\left(\frac{itn}{2}\left(1+\frac{i\vartheta}{\pi}\right)\right)\right]\,,\label{eq:f11}
\end{equation}
where the normalisation $\mathcal{N}$ guarantees $f_{11}(\pm\infty)=1$. Finally we have
\begin{equation}
F_{\text{min}}^{\mathcal{T}|(\beta1),(\gamma1)}(\vartheta,n)=\mathcal{N}\exp\left[\int_{0}^{\infty}\frac{\mathrm{d}t}{t\sinh nt}g_{\beta,\gamma}(t)\sin^{2}\left(\frac{it}{2}\left(n+\frac{i\vartheta}{\pi}\right)\right)\right]\,.
\end{equation}

The minimal form factors are the building blocks to obtain
all form factors with particle number $k\geq2$, simplifying the solution
of the bootstrap equations. We recall that the zero and the one-particle FFs must be obtained by other means. 
In the absence of bound states,
the two-particle form factors  (which are the most relevant FFs) for the branch-point twist field, satisfying
also the kinematic poles axioms, have been determined as \cite{ccd-08}
\begin{equation}
F_{2}^{\mathcal{T}|(\beta k),(\bar{\beta}j)}(\vartheta,n)=\frac{\langle\mathcal{T}_{n}\rangle\sin\frac{\pi}{n}}{2n\sinh\left(\frac{i\pi\left(2(j-k)-1\right)+\vartheta}{2n}\right)\sinh\left(\frac{i\pi\left(2(k-j)-1\right)-\vartheta}{2n}\right)}\frac{F_{\text{min}}^{\mathcal{T}|(\beta k),(\bar{\beta}j)}(\vartheta,n)}{F_{\text{min}}^{\mathcal{T}|(\beta1),(\bar{\beta}1)}(i\pi,n)}\,,\label{eq:F2TwistField}
\end{equation}
where $\langle\mathcal{T}_{n}\rangle=F_{0}^{\mathcal{T}}$ is the
vacuum expectation value (VEV) of ${\cal T}$. Furthermore, relativistic
invariance implies that $F_{2}^{\mathcal{T}|(\beta k),(\gamma j)}(\vartheta_{1},\vartheta_{2},n)$
depends only on the rapidity difference $\vartheta_{1}-\vartheta_{2},$
justifying writing $F_{2}^{\mathcal{T}|(\beta k),(\gamma j)}(\vartheta_{1}-\vartheta_{2},n)$
or simply $F_{2}^{\mathcal{T}|(\beta k),(\gamma j)}(\vartheta_{12},n)$.
Because of the replica symmetry, for $\hat{\mathcal{T}}$ we have
\begin{equation}
F_{2}^{\mathcal{T}|(\beta k),(\gamma j)}(\vartheta,n)=F_{2}^{\hat{\mathcal{T}}|(\beta n-j),(\gamma n-k)}(\vartheta,n)\,.\label{FFtil}
\end{equation}

\subsection{Branch-point twist field form factors in the Dirac field theory}

The free Dirac theory has a massive fermion particle (denoted by $+$)
and anti-particle (denoted by $-$) and the simple S-matrix with elements
\begin{equation}
S_{\text{\ensuremath{\pm\pm}}}(\vartheta)=S_{\text{\ensuremath{\pm\mp}}}(\vartheta)=-1.
\end{equation}
Consequently we have 
\begin{equation}
F_{\text{min}}^{\mathcal{T}_{D}|(\pm1),(\mp1)}(\vartheta,n)=F_{\text{min}}^{\mathcal{T}_{D}|1,1}(\vartheta,n)=-i\sinh\frac{\vartheta}{2n}\,.
\end{equation}
For this model, the FFs of the branch-point twist fields are only
non-vanishing for even particle number \cite{ccd-08,cd-09}. Moreover,
the FFs for any even $n$ can be written as a Pfaffain of the two-particle
FF \cite{cd-09b}.

The FFs of the branch-point twist field is only non-vanishing for
neutral states containing an equal number of particles and anti-particles.
In the Dirac theory, in particular, we have \cite{ccd-08,bcd-16}
\begin{equation}
\begin{split}F_{2}^{\mathcal{T}_{D}|(+j),(-k)}(\vartheta,n)=\frac{\langle\mathcal{T}_{D,n}\rangle\sin\frac{\pi}{n}}{2n\sinh\left(\frac{i\pi\left(2(j-k)-1\right)+\vartheta}{2n}\right)\sinh\left(\frac{i\pi\left(2(k-j)-1\right)-\vartheta}{2n}\right)}\frac{F_{\text{min}}^{\mathcal{T}_{D}|j,k}(\vartheta,n)}{F_{\text{min}}^{\mathcal{T}_{D}|1,1}(i\pi,n)}\,,\\
F_{2}^{\mathcal{T}_{D}|(-j),(+k)}(\vartheta,n)=\frac{\langle\mathcal{T}_{D,n}\rangle\sin\frac{\pi}{n}}{2n\sinh\left(\frac{i\pi\left(2(j-k)-1\right)+\vartheta}{2n}\right)\sinh\left(\frac{i\pi\left(2(k-j)-1\right)-\vartheta}{2n}\right)}\frac{F_{\text{min}}^{\mathcal{T}_{D}|j,k}(\vartheta,n)}{F_{\text{min}}^{\mathcal{T}|_{D}1,1}(i\pi,n)}\,,
\end{split}
\label{eq:F2TwistFieldDirac}
\end{equation}
that is $F_{2}^{\mathcal{T}_{D}|(+j),(-k)}=F_{2}^{\mathcal{T}_{D}|(-j),(+k)}$.
$\mathcal{T}_{D}$ refers to the branch-point twist field of the Dirac
theory and accordingly $\langle\mathcal{T}_{D,n}\rangle$ denotes
the VEV of this field in the $n$-sheeted theory.

\subsection{Branch-point twist field form factors in the free complex boson theory}

Also, the free complex boson theory has a particle ($+$) and an antiparticle
($-$). The S-matrix is identically 1 between and within each species,
so 
\begin{equation}
S_{\text{\ensuremath{\pm\pm}}}(\vartheta)=S_{\text{\ensuremath{\pm\mp}}}(\vartheta)=1,
\end{equation}
where $\pm$ refers to particles and anti-particles. The minimal form
factor is also unity
\begin{equation}
F_{\text{min}}^{\mathcal{T}_{B}|(\pm1),(\pm1)}(\vartheta,n)=F_{\text{min}}^{\mathcal{T}_{B}|1,1}(\vartheta,n)=1\,,
\end{equation}
for any $n$.

The FFs of the branch-point twist field is only non-vanishing for
neutral states containing an equal number of particles and anti-particles.
In particular, we have the simple expressions \cite{bcd-16}
\begin{equation}
\begin{split}F_{2}^{\mathcal{T}_{B}|(+j),(-k)}(\vartheta,n)=\frac{\langle\mathcal{T}_{B,n}\rangle\sin\frac{\pi}{n}}{2n\sinh\left(\frac{i\pi\left(2(j-k)-1\right)+\vartheta}{2n}\right)\sinh\left(\frac{i\pi\left(2(k-j)-1\right)-\vartheta}{2n}\right)}\,,\\
F_{2}^{\mathcal{T}_{B}|(-j),(+k)}(\vartheta,n)=\frac{\langle\mathcal{T}_{B,n}\rangle\sin\frac{\pi}{n}}{2n\sinh\left(\frac{i\pi\left(2(j-k)-1\right)+\vartheta}{2n}\right)\sinh\left(\frac{i\pi\left(2(k-j)-1\right)-\vartheta}{2n}\right)}\,.
\end{split}
\label{eq:F2TwistFieldBoson}
\end{equation}
In this case we find again that $F_{2}^{\mathcal{T}_{B}|(+j),(-k)}=F_{2}^{\mathcal{T}_{B}|(-j),(+k)}$.
$\mathcal{T}_{B}$ now refers to the branch-point twist field of the
free massive complex boson theory and accordingly $\langle\mathcal{T}_{B,n}\rangle$
denotes the VEV of this field in the $n$-sheeted theory.

\subsection{The form factors of the stress energy tensor $\Theta$ in free theories}

Later on we will need also the FFs of the $\Theta$ field and so we
report here some basic details about them. In both the free massive
Dirac and the free massive complex boson theory this field has only
a non-vanishing 2-particle FF, which reads 
\begin{equation}
F_{D,2}^{\Theta|(\pm),(\mp)}(\theta)=-i2\pi m^{2}\sinh\frac{\vartheta}{2},
\end{equation}
for the 1-copy Dirac and 
\begin{equation}
F_{B,2}^{\Theta|(\pm),(\mp)}(\theta)=2\pi m^{2},
\end{equation}
for the 1-copy complex boson theory, where $m$ is mass of the fermion
and boson. In the $n$-copy theories, the FF of this field behaves
in an additive way, that is \cite{ccd-08} 
\begin{equation}
F_{D,2}^{\Theta|(\pm j),(\mp k)}(\theta,n)=\begin{cases}
-i2\pi m^{2}\sinh\frac{\vartheta}{2} & j=k\\
0 & j\neq k
\end{cases}
\end{equation}
and 
\begin{equation}
F_{B,2}^{\Theta|(\pm j),(\mp k)}(\theta,n)=\begin{cases}
2\pi m^{2} & j=k\\
0 & j\neq k\,.
\end{cases}
\end{equation}

\section{Form factors of the composite U(1) branch-point twist field in the
Dirac theory \label{secDirac}}

\label{sec:U1DiracTwistFields}

The bootstrap equations for the FFs of the branch-point twist field
can be naturally modified to obtain the corresponding quantities of
the composite branch-point twist fields \cite{Z2IsingShg}. We define
the semi-local (or mutual locality) index $e^{i\alpha}$ of an operator
$O$ with respect to another field $\phi$ via the condition
\begin{equation}
O(x,t)\phi(y,t')=e^{i\alpha}\phi(y,t')O(x,t),\label{eq:LocalityDefBIS}
\end{equation}
or, when using the radial quantisation picture, 
\begin{equation}
O(e^{-i2\pi}z,e^{i2\pi}\bar{z})\phi(0,0)=e^{i\alpha}O(z,\bar{z})\phi(0,0)\,.\label{eq:LocalityDef}
\end{equation}
Mutually local operators correspond to $e^{i\alpha}=1$, while fields
with $e^{i\alpha}\neq1$ are called mutually semi-local. 
In Ref. \cite{Z2IsingShg} it was argued that it is natural to assume that the flux
phase $e^{i\alpha}$ can be related with the mutual locality index
appearing in the bootstrap equation. This assumption can be based
on the intuitive picture associated with the insertion of the Aharonov-Bohm
flux on one of the Riemann sheets. In this picture, the flux is carried
by the particles of the theory, but Eq. (\ref{eq:LocalityDef}) is
just an equivalent rephrasing of this idea when $\phi$ is an interpolating
field associated with creation or annihilation of particles.
These ideas can be equivalently rephrased by writing the exchange relations
for the composite branch-point twist fields of the $n$-copy theory as
\begin{equation}
\begin{split}\Psi_{p,i}(y)\mathcal{T}^{\alpha}(x)=e^{ip\alpha/n}\mathcal{T}^{\alpha}(x)\Psi_{p,i+1}(y) & \qquad x<y,\\
\Psi_{p,i}(y)\mathcal{T}^{\alpha}(x)=\mathcal{T}^{\alpha}(x)\Psi_{p,i}(y) & \qquad x>y,
\end{split}
\end{equation}
and similarly for $\tilde{\mathcal{T}}$
\begin{equation}
\begin{split}\Psi_{p,i}(y)\mathcal{\tilde{\mathcal{T}}}^{\alpha}(x)=e^{-ip\alpha/n}\mathcal{\tilde{\mathcal{T}}}^{\alpha}(x)\Psi_{p,i-1}(y) & \qquad x{>y},\\
\Psi_{p,i}(y)\mathcal{\tilde{\mathcal{T}}}^{\alpha}(x)=\tilde{\mathcal{T}}^{\alpha}(x)\Psi_{p,i}(y) & \qquad x<y,
\end{split}
\end{equation}
where $\Psi_{p,i}(x)$ is a generic quantum field living on the  $i$th replica and possessing a definite  $U(1)$ charge  $p\in\mathbb{Z}$.
Our choice for $\alpha/n$  is dictateted by the requirement, that the total
phase picked up by the particle when turning around the entire
Riemann surface has to be  $e^{i\alpha}$. Specialising now the bootstrap equations of a semi-local $U(1)$ composite
branch-point twist field $\mathcal{T}^{\alpha}$ to the Dirac theory, we can write 
\begin{eqnarray}
 &  & F_{k}^{\mathcal{T}_{D}^{\alpha}|\underline{(\beta\mu)}}(\underline{\vartheta})=S_{(\beta_{i}\mu_{i}),(\beta_{i+1}\mu_{i+1})}(\vartheta_{i,i+1})F_{k}^{\mathcal{T}_{D}^{\alpha}|...(\beta_{i+1}\mu_{i+1}),(\beta_{i}\mu_{i})...}(\ldots\vartheta_{i+1},\vartheta_{i},\ldots),\label{eq:FFAxiom1SemiLocalGeneral}\\
 &  & F_{k}^{\mathcal{T}_{D}^{\alpha}|\underline{(\beta\mu)}}(\vartheta_{1}+2\pi i,\vartheta_{2},\ldots,\vartheta_{k})=e^{i\alpha\beta_{1}/n}F_{k}^{\mathcal{T}_{D}^{\alpha}|(\beta_{2}\mu_{2}),...,(\beta_{k}\mu_{k}),(\beta_{1}\hat{\mu}_{1})}(\vartheta_{2},\ldots,\vartheta_{n},\vartheta_{1}),\label{eq:FFAxiom2SemiLocalGeneral}\\
 &  & -i\underset{\vartheta_{0}'=\vartheta_{0}+i\pi}{{\rm Res}}F_{k+2}^{\mathcal{T}_{D}^{\alpha}|(\beta \mu),(\bar{\beta} \mu),\underline{(\beta\mu)}}(\vartheta_{0}',\vartheta_{0},\underline{\vartheta})=F_{k}^{\mathcal{T}_{D}^{\alpha}|\underline{(\beta\mu)}}(\underline{\vartheta}),\label{eq:FFAxiom3SemiLocalGeneral}\\
 &  & -i\underset{\vartheta_{0}'=\vartheta_{0}+i\pi}{{\rm Res}}F_{k+2}^{\mathcal{T}_{D}^{\alpha}|(\beta \mu),(\bar{\beta} \hat{\mu}),\underline{(\alpha\mu)}}(\vartheta_{0}',\vartheta_{0},\underline{\vartheta})=-e^{i\alpha\beta/n}\prod_{i=1}^{k}S_{(\beta \hat{\mu}),(\beta_{i}\mu_{i})}(\vartheta_{0i})F_{k}^{\mathcal{T}_{D}^{\alpha}|\underline{(\beta\mu)}}(\underline{\vartheta}),\nonumber 
\end{eqnarray}
where $\beta$ and $\beta_{i}=\pm1$ (sometimes simply shortened in
$\pm$). $\mathcal{T}_{D}^{\alpha}$ denotes the composite $U(1)$
branch-point twist field in the free massive Dirac QFT. To avoid confusion,
the standard $U(1)$ twist field in the Dirac theory is denoted by
$\mathcal{V}_{D}^{\alpha}$. This field is also referred to as the
vertex operator, but based on its monodromy properties it can be equivalently
called a $U(1)$-twist field. In both cases, the superscript $\alpha$
corresponds to the inserted flux.

The FFs of the composite $U(1)$ branch-point twist fields are non-zero
for even number of particles and charge-neutral configurations. This
is also true for $\mathcal{V}_{D}^{\alpha}$ which can be obtained
from the above equations for $n=1$. It is instructive to write down
its 2-particle FFs, known from earlier investigations \cite{LeClairU1,KarowskiU1,SwiecaU1}, 
\begin{equation}
F_{2}^{\mathcal{V}_{D}^{\alpha}|(\pm),(\mp)}(\vartheta,\alpha)=\pm\frac{i2\sin\frac{\alpha}{2}e^{\pm\frac{\vartheta\alpha}{2\pi}}\sinh\frac{\vartheta}{2}}{\sinh\vartheta}\,.\label{eq:F2U(1)Dirac}
\end{equation}
Having obtained the defining equations for the form factors, following
the logic of section \ref{sec:ConventionalTwistFields}, we can write
\begin{equation}
\begin{split}F_{\text{min}}^{\mathcal{T}_{D}^{\alpha}|(+k),(-j)}(\vartheta,n)=F_{\text{min}}^{\mathcal{T}_{D}^{\alpha}|(-j),(+k)}(-\vartheta,n)S_{k,j}(\vartheta)=e^{+i\alpha/n}F_{\text{min}}^{\mathcal{T}_{D}^{\alpha}|(-j),(+k+1)}(2\pi i-\vartheta,n)\,,\\
F_{\text{min}}^{\mathcal{T}_{D}^{\alpha}|(-k),(+j)}(\vartheta,n)=F_{\text{min}}^{\mathcal{T}_{D}^{\alpha}|(+j),(-k)}(-\vartheta,n)S_{k,j}(\vartheta)=e^{-i\alpha/n}F_{\text{min}}^{\mathcal{T}_{D}^{\alpha}|(+j),(k+1)}(2\pi i-\vartheta,n)\,,
\end{split}
\end{equation}
for the minimal form factor $F_{\text{min}}^{\mathcal{T}_{D}^{\alpha}}$
of the composite branch-point twist field. From this we find 
\begin{equation}
\begin{split}F_{\text{min}}^{\mathcal{T}_{D}^{\alpha}|(\pm i),(\mp i+k)}(\vartheta,n)=F_{\text{min}}^{\mathcal{T}_{D}^{\alpha}|(\pm j),(\mp j+k)}(\vartheta,n) & \ \quad\forall i,j,k,\\
F_{\text{min}}^{\mathcal{T}_{D}^{\alpha}|(\mp1),(\pm j)}(\vartheta,n)=e{}^{\mp i\alpha(j-1)/n}F_{\text{min}}^{\mathcal{T}_{D}^{\alpha}|(\pm1),(\mp1)}(2\pi i(j-1)-\vartheta,n) & \ \quad\forall j\neq1,
\end{split}
\label{eq:FMinAxiomsSemiLocalU1}
\end{equation}
and finally we get 
\begin{equation}
F_{\text{min}}^{\mathcal{T}_{D}^{\alpha}|(\pm j),(\mp k)}(\vartheta,n)=\begin{cases}
e^{\pm i\alpha(k-j)/n}F_{\text{min}}^{\mathcal{T}_{D}^{\alpha}|(\mp1),(\pm1)}(2\pi i(k-j)-\vartheta,n) & \text{ if }k>j,\\
e^{\mp i\alpha(j-k)/n}F_{\text{min}}^{\mathcal{T}_{D}^{\alpha}|(\pm1),(\mp1)}(2\pi i(j-k)+\vartheta,n) & \text{otherwise.}
\end{cases}\label{eq:FDMinBasicProp}
\end{equation}
Akin to the previous case, the only independent quantity is $F_{\text{min}}^{\mathcal{T}_{D}^{\alpha}|(\pm1),(\mp1)}(\vartheta,n)$.
We exploit Eq. (\ref{eq:FMinAxiomsSemiLocalU1}) to write 
\begin{equation}
F_{\text{min}}^{\mathcal{T}_{D}^{\alpha}|(\pm1),(\mp1)}(\vartheta,n)=-F_{\text{min}}^{\mathcal{T}_{D}^{\alpha}|(\mp1),(\pm1)}(-\vartheta,n)=e^{\pm i\alpha}F_{\text{min}}^{\mathcal{T}_{D}^{\alpha}|(\mp1),(\pm1)}(-\vartheta+2\pi in,n)\,.
\end{equation}
The solution of $F_{\text{min}}^{\mathcal{T}_{D}^{\alpha}|(\pm1),(\mp1)}$
can be obtained by introducing $f_{(\pm1),(\mp1)}^{\mathcal{T}_{D}^{\alpha}}(\vartheta)$
as 
\begin{equation}
F_{\text{min}}^{\mathcal{T}_{D}^{\alpha}|(\pm1),(\mp1)}(\vartheta,n)=f_{(\pm1),(\mp1)}^{\mathcal{T}_{D}^{\alpha}}(\vartheta/n)\,,
\end{equation}
that satisfies 
\begin{equation}
f_{(\pm1),(\mp1)}^{\mathcal{T}_{D}^{\alpha}}(\vartheta)=-f_{(\pm1),(\mp1)}^{\mathcal{T}_{D}^{\alpha}}(-\vartheta)=e^{\pm i\alpha}f_{(\pm1),(\mp1)}^{\mathcal{T}_{D}^{\alpha}}(-\vartheta+2\pi i)\,.\label{f11axiomSemiLocalU1}
\end{equation}
Luckily, $f_{(\pm1),(\mp1)}^{\mathcal{T}_{D}^{\alpha}}$ can be easily
obtained from $f_{11}$ by multiplying the latter by an appropriately
chosen CDD factor, $f_{\text{CDD}}^{\mathcal{T}_{D}^{\alpha}|(\pm1),(\mp1)}$.
Such a factor must obey 
\begin{equation}
f_{\text{CDD}}^{\mathcal{T}_{D}^{\alpha}|(\pm1),(\mp1)}(\vartheta)=f_{\text{CDD}}^{\mathcal{T}_{D}^{\alpha}|(\pm1),(\mp1)}(-\vartheta)=e^{\pm i\alpha}f_{\text{CDD}}^{\mathcal{T}_{D}^{\alpha}|(\pm1),(\mp1)}(-\vartheta+2\pi i),\label{eq:fCDDAxiom}
\end{equation}
guaranteeing that $f_{(\pm1),(\mp1)}^{\mathcal{T}_{D}^{\alpha}}(\vartheta)=f_{\text{CDD}}^{\mathcal{T}_{D}^{\alpha}|(\pm1),(\mp1)}(\vartheta)f_{11}(\vartheta)$
satisfies Eq. (\ref{f11axiomSemiLocalU1}). The correct choice for
$f_{\text{CDD}}$ turns out to be 
\begin{equation}
f_{\text{CDD}}^{\mathcal{T}_{D}^{\alpha}|(\pm1),(\mp1)}(\vartheta)=e^{\pm\frac{\vartheta\alpha}{2\pi}}\,,\label{eq:fCDDSemiLocalU1}
\end{equation}
and hence the minimal FF is 
\begin{equation}
F_{\text{min}}^{\mathcal{T}_{D}^{\alpha}|(\pm1),(\mp1)}(\vartheta,n,\alpha)=-ie^{\pm\frac{\vartheta\alpha}{2\pi n}}\sinh\frac{\vartheta}{2n}\,.
\end{equation}
It is easy to check that the ansatz (\ref{eq:fCDDSemiLocalU1}) satisfies
Eq. (\ref{eq:fCDDAxiom}), and any further CDD ambiguity is excluded
by exploiting the $\Delta$-theorem checks performed in the next sections.

With the minimal form factor at hand, we can write the 2-particle
form factor as a product 
\begin{equation}
F_{2}^{\mathcal{T}_{D}^{\alpha}|(\pm j),(\mp k)}(\vartheta,n,\alpha)=P_{D}^{(\pm j),(\mp k)}(\vartheta,n,\alpha)F_{\text{min}}^{\mathcal{T}_{D}^{\alpha}|(\pm j),(\mp k)}(\vartheta,n,\alpha)\,,\label{eq:F2TU1Product}
\end{equation}
where the function $P_{D}^{(\pm j),(\mp k)}$ accounts for the pole
structure in Eq. \eqref{eq:FFAxiom3SemiLocalGeneral}. Using Eqs.
\eqref{eq:FFAxiom1SemiLocalGeneral} and \eqref{eq:FFAxiom2SemiLocalGeneral},
one finds that 
\begin{equation}
P_{D}^{(\pm j),(\mp k)}(\vartheta,n,\alpha)=\begin{cases}
P_{D}^{(\mp1j),(\pm1)}(2\pi i(k-j)-\vartheta,n,\alpha) & \text{ if }k>j,\\
P_{D}^{(\pm1j),(\mp1)}(2\pi i(j-k)+\vartheta,n,\alpha) & \text{otherwise.}
\end{cases}\label{eq:PBasicProp}
\end{equation}
The only independent quantity is $P^{(+1),(-1)}$ satisfying 
\begin{equation}
P_{D}^{(+1),(-1)}(\vartheta,n,\alpha)=P_{D}^{(-1),(+1)}(-\vartheta,n,\alpha),\label{eq:ParitypropP}
\end{equation}
together with the important property 
\begin{equation}
\begin{split}P_{D}^{(+1),(-1)}(2\pi ni+\vartheta,n,\alpha)=P_{D}^{(-1),(+1)}(-\vartheta,n,\alpha),\\
P_{D}^{(-1),(+1)}(2\pi ni+\vartheta,n,\alpha)=P_{D}^{(+1),(-1)}(-\vartheta,n,\alpha).
\end{split}
\label{eq:CyclicProeprtyP}
\end{equation}
Combining Eq. \eqref{eq:FFAxiom3SemiLocalGeneral} with the minimal form factor, we have 
\begin{equation}
\begin{split}-i\underset{x=0}{{\rm Res}}\,P_{D}^{(\pm1),(\mp1)}(x+i\pi,n,\alpha)\sinh\frac{i\pi}{2n}e^{\pm i\alpha/(2n)}= & 1,\\
-i\underset{x=0}{{\rm Res}}\,P_{D}^{(\pm1),(\mp1)}(x-i\pi,n,\alpha)\sinh\frac{i\pi}{2n}e^{\mp i\alpha/(2n)}e^{\pm i\alpha/(n)}= & -e^{\pm i\alpha/n}\,.
\end{split}
\label{eq:Axiom3P}
\end{equation}
We can easily check that for $\alpha=0$, i.e., the case of the conventional
branch-point twist field when $P_{D}^{(+1),(-1)}$ equals $P_{D}^{(-1),(+1)}$,
the function 
\begin{equation}
P(\vartheta,n,0)=\frac{\sin\frac{\pi}{n}}{2n\sinh\frac{i\pi-\vartheta}{2n}\sinh\frac{i\pi+\text{\ensuremath{\vartheta}}}{2n}\sinh\frac{i\pi}{2n}}
\end{equation}
solves the above equations. When $\alpha\neq0$, we have to account
for the complex phase factors but still respecting the cyclic property
\eqref{eq:CyclicProeprtyP} of $P_{D}$. This can be achieved multiplying
$P(\vartheta,n,0)$ by $2\pi n$ periodic functions along the imaginary
axis, whose values at $i\pi$ and $-i\pi$ are $e^{\mp i\alpha/(2n)}$
and $e^{\pm i\alpha/(2n)}$, respectively. The simplest choice for
such a function is 
\begin{equation}
p_{D}^{\pm}(\vartheta,n,\alpha)=\cos\frac{\alpha}{2n}\mp\sin\frac{\alpha}{2n}\frac{\sinh\frac{\vartheta}{n}}{\sin\frac{\pi}{n}}\,.
\end{equation}
$P_{D}^{(\pm1),(\mp1)}(\vartheta,n,\alpha)=p_{D}^{\pm}(\vartheta,n,\vartheta)P(\vartheta,n,0)$
satisfies the parity and cyclic property Eqs. \eqref{eq:ParitypropP}
and \eqref{eq:CyclicProeprtyP} and the equivalent of \eqref{eq:FFAxiom3SemiLocalGeneral},
i.e., \eqref{eq:Axiom3P}. Furthermore, $p_{D}^{\pm}(\vartheta,n,\alpha)$
does not introduce any additional poles on the physical sheet.

Using thus \eqref{eq:FDMinBasicProp} and the product formula \eqref{eq:F2TU1Product}
for the 2-particle FF, all the defining axioms \eqref{eq:FFAxiom1SemiLocalGeneral},
\eqref{eq:FFAxiom2SemiLocalGeneral}, \eqref{eq:FFAxiom3SemiLocalGeneral}
are satisfied. Unfortunately this solution is not the correct one
corresponding to the composite $U(1)$ branch-point twist field. Although
for $\alpha=0$ the FF of the conventional twist field is recovered,
for $n\rightarrow1$ the solution is also expected to simplify to
the conventional $U(1)$ twist field form factor, which is not the
case. Based on this quantity the right choice for $p^{\pm}$ can be
determined. The correct solution for $P_{D}^{(\pm1),(\mp1)}(\vartheta,n,\alpha)$
reads 
\begin{equation}
P_{D}^{(\pm1),(\mp1)}(\vartheta,n,\vartheta)=\frac{\sin\frac{\pi}{n}}{2n\sinh\frac{i\pi-\vartheta}{2n}\sinh\frac{i\pi+\text{\ensuremath{\vartheta}}}{2n}\sinh\frac{i\pi}{2n}}\left(\cos\frac{\alpha}{2n}\pm\sin\frac{\alpha}{2n}\frac{\sin^{2}\frac{\pi}{2n}\sinh\frac{\vartheta}{n}}{\sin\frac{\pi}{n}\sinh^{2}\frac{\vartheta}{2n}}\right)\,,\label{eq:PSolution}
\end{equation}
in which the correct choice for $p_{D}^{\pm}(\vartheta,n,\alpha)$
\begin{equation}
p_{D}^{\pm}(\vartheta,n,\alpha)=\cos\frac{\alpha}{2n}\pm\sin\frac{\alpha}{2n}\frac{\sin^{2}\frac{\pi}{2n}\sinh\frac{\vartheta}{n}}{\sin\frac{\pi}{n}\sinh^{2}\frac{\vartheta}{2n}}
\end{equation}
does not introduce additional poles on the physical sheet (similarly
to our previous naive choice) and for large rapidities it tends to
a constant. Using the formula \eqref{eq:PSolution} to construct the
2-particle form factors via \eqref{eq:F2TU1Product} and \eqref{eq:PBasicProp}
all the axioms \eqref{eq:FFAxiom1SemiLocalGeneral}, \eqref{eq:FFAxiom2SemiLocalGeneral},
\eqref{eq:FFAxiom3SemiLocalGeneral} are satisfied and the $\alpha\rightarrow0$
and $n\rightarrow1$ limits yield the desired results. i.e., the conventional
branch-point \eqref{eq:F2TwistFieldDirac} and the standard $U(1)$
twist field FFs \eqref{eq:F2U(1)Dirac}, which one can easily check.
Performing algebraic simplifications, the full two-particle FF can be
therefore written as 
\begin{equation}
F_{2}^{\mathcal{T}_{D}^{\alpha}|(\pm1),(\mp1)}(\vartheta,n,\alpha)=\frac{-i\langle\mathcal{T}_{D,n}^{\alpha}\rangle\sin\frac{\pi}{n}}{2n\sinh\frac{i\pi-\vartheta}{2n}\sinh\frac{i\pi+\text{\ensuremath{\vartheta}}}{2n}}\left(\frac{\cos\frac{\alpha}{2n}\sinh\frac{\vartheta}{2n}}{\sin\frac{\pi}{2n}}\pm\frac{\sin\frac{\alpha}{2n}\cosh\frac{\vartheta}{2n}}{\cos\frac{\pi}{2n}}\right)e^{\pm\frac{\vartheta\alpha}{2\pi n}}\,,\label{eq:F2DiracSolution11}
\end{equation}
and, using Eq. \eqref{eq:FDMinBasicProp}, we arrive at 
\begin{equation}
F_{2}^{\mathcal{T}_{D}^{\alpha}|(\pm j),(\mp k)}(\vartheta,n)=\begin{cases}
e^{\pm i\alpha(k-j)/n}F_{2}^{\mathcal{T}_{D}^{\alpha}|(\mp1),(\pm1)}(2\pi i(k-j)-\vartheta,n) & \text{ if }k>j,\\
e^{\mp i\alpha(j-k)/n}F_{2}^{\mathcal{T}_{D}^{\alpha}|(\pm1),(\mp1)}(2\pi i(j-k)+\vartheta,n) & \text{otherwise.}
\end{cases}\label{eq:FD2Full}
\end{equation}
In all the above treatment $\alpha$ is within the range $[-\pi,\pi]$,
otherwise descendant fields are obtained.

\subsection{$\Delta$-theorem test and higher-particle FFs}

The validity of the solution can be checked by the $\Delta$-theorem, which for the Dirac field can be written as 
\begin{equation}
-\frac{n}{32\pi^{2}m^{2}\langle\mathcal{T}_{D}^{\alpha}\rangle}\int\mathrm{d}\vartheta\frac{F_{D,2}^{\Theta|1,1}\left(\vartheta\right)\left(F_{2}^{\mathcal{T}_{D}^{\alpha}|(\pm1),(\mp1)}(\vartheta,n)^{*}+F_{2}^{\mathcal{T}_{D}^{\alpha}|(\mp1),(\pm1)}(\vartheta,n)^{*}\right)}{\cosh^{2}\left(\vartheta/2\right)}\,,
\end{equation}
where we (i) used that in the Dirac theory the field $\Theta$ has only
2-particle FFs (ii) we performed a changed of variables (iiI) we did a final integration in the $k=2$ term of Eq. \eqref{eq:DeltaTheoreM3}. 
The chiral conformal dimensions obtained from the appropriate integrals of the 2-particle FFs gives the exact UV result \cite{gs-18}
\begin{equation}
\Delta_{n}^{\mathcal{T}_{D}^{\alpha}}=\frac{1}{24}\left(n-n^{-1}\right)+\frac{\left(\alpha/(2\pi)\right)^{2}}{2n}\,.
\end{equation}

It is also easy to obtain the FFs of higher particle numbers in
terms of a Pfaffian. Assuming the ordering of the replica index $\mu_{1}\geq\mu_{2}\geq...\geq\mu_{2k}$,
one has 
\begin{equation}
F_{2k}^{\mathcal{T}_{D}^{\alpha}|\underline{(\beta\mu)}}(\underline{\vartheta})=\langle\mathcal{T}_{D}^{\alpha}\rangle\text{Pf}(W)\,,\label{eq:Pfaffain}
\end{equation}
where $W$ is a $2k\times2k$ anti-symmetric matrix with entries 
\begin{equation}
W_{lm}=\begin{cases}
\frac{F_{2}^{\mathcal{T}_{D}^{\alpha}|(\beta_{l}\mu_{l}),(\beta_{m}\mu_{m})}(\vartheta_{l}-\vartheta_{m},n)}{\langle\mathcal{T}_{D}^{\alpha}\rangle} & m>l,\\
\left(-1\right)^{\delta_{\mu_{l},\mu_{m}}+1}\frac{F_{2}^{\mathcal{T}_{D}^{\alpha}|(\beta_{l}\mu_{l}),(\beta_{m}\mu_{m})}(\vartheta_{l}-\vartheta_{m},n)}{\langle\mathcal{T}_{D}^{\alpha}\rangle} & m<l\,,
\end{cases}
\end{equation}
where $F_{2}^{\mathcal{T}_{D}^{\alpha}|(\beta_{l}\mu_{l}),(\beta_{m}\mu_{m})}$
is zero if $\beta_{l}=\beta_{m}=\pm1$, a manifestation of neutrality
condition of the composite $U(1)$ branch point twist field. 

Finally, we note that the FFs of the other field $\tilde{\mathcal{T}}_{D}^{\alpha}$
can be easily obtained from that of $\mathcal{T}_{D}^{\alpha}$. 
As an example, for the 2-particle case we have
\begin{equation}
F_{2}^{\tilde{\mathcal{T}}_{D}^{\alpha}|(\pm j),(\mp k)}(\vartheta,n,\alpha)=F_{2}^{\mathcal{T}_{D}^{-\alpha}|(\pm n-j),(\mp n-k)}(\vartheta,n,-\alpha)\,.
\end{equation}

\section{Form factors of the composite U(1) branch-point twist field in the
complex boson theory \label{secBoson}}

\label{sec:U1BosonTwistFields}

In this section we treat the case of the free complex boson for which the bootstrap equations of a semi-local U(1) composite branch-point
twist field are
\begin{eqnarray}
 &  & F_{k}^{\mathcal{T}_{B}^{\alpha}|\underline{(\beta\mu)}}(\underline{\vartheta})=F_{k}^{\mathcal{T}_{B}^{\alpha}|...(\beta_{i+1}\mu_{i+1}),(\beta_{i}\mu_{i})..}(\ldots\vartheta_{i+1},\vartheta_{i},\ldots),\label{eq:FFAxiom1SemiLocalGeneralBoson}\\
 &  & F_{k}^{\mathcal{T}_{B}^{\alpha}|\underline{(\beta\mu)}}(\vartheta_{1}+2\pi i,\vartheta_{2},\ldots,\vartheta_{k})=e^{i\alpha\beta_{1}/n}F_{k}^{\mathcal{T}_{B}^{\alpha}|(\beta_{2}\mu_{2}),...,(\beta_{k}\mu_{k}),(\beta_{1}\hat{\mu}_{1})}(\vartheta_{2},\ldots,\vartheta_{n},\vartheta_{1}),\label{eq:FFAxiom2SemiLocalGeneralBoson}\\
 &  & -i\underset{\vartheta_{0}'=\vartheta_{0}+i\pi}{{\rm Res}}F_{k+2}^{\mathcal{T}_{B}^{\alpha}|(\beta \mu),(\bar{\beta} \mu),\underline{(\beta\mu)}}(\vartheta_{0}',\vartheta_{0},\underline{\vartheta})=F_{k}^{\mathcal{T}_{B}^{\alpha}|\underline{(\beta\mu)}}(\underline{\vartheta}),\label{eq:FFAxiom3SemiLocalGeneralBoson}\\
 &  & -i\underset{\vartheta_{0}'=\vartheta_{0}+i\pi}{{\rm Res}}F_{k+2}^{\mathcal{T}_{B}^{\alpha}|(\beta \mu),(\bar{\beta} \hat{\mu}),\underline{(\beta\mu)}}(\vartheta_{0}',\vartheta_{0},\underline{\vartheta})=-e^{i\alpha\beta/n}F_{k}^{\mathcal{T}_{B}^{\alpha}|\underline{(\beta\mu)}}(\underline{\vartheta}),\nonumber 
\end{eqnarray}
since the S-matrix is identically $+1$ within and between replicas.
Again $\beta$ and $\beta_{i}=\pm1$ often abbreviated as merely $\pm$
and the FFs of the composite $U(1)$ branch-point twist fields are
non-zero for even number of particles and charge-neutral configurations.
Following the logic of section \ref{sec:U1DiracTwistFields}, we can
write the  equations 
\begin{equation}
\begin{split}F_{\text{min}}^{\mathcal{T}_{B}^{\alpha}|(+k),(-j)}(\vartheta,n)=F_{\text{min}}^{\mathcal{T}_{B}^{\alpha}|(-j),(+k)}(-\vartheta,n)=e^{+i\alpha/n}F_{\text{min}}^{\mathcal{T}_{B}^{\alpha}|(-j),(+k+1)}(2\pi i-\vartheta,n)\,,\\
F_{\text{min}}^{\mathcal{T}_{B}^{\alpha}|(-k),(+j)}(\vartheta,n)=F_{\text{min}}^{\mathcal{T}_{B}^{\alpha}|(+j),(-k)}(-\vartheta,n)=e^{-i\alpha/n}F_{\text{min}}^{\mathcal{T}_{B}^{\alpha}|(+j),(k+1)}(2\pi i-\vartheta,n)\,,
\end{split}
\end{equation}
for the minimal form factor $F_{\text{min}}^{\mathcal{T}_{B}^{\alpha}}$ of the composite branch-point twist field. 
From this we find 
\begin{equation}
\begin{split}F_{\text{min}}^{\mathcal{T}_{B}^{\alpha}|(\pm i),(\mp i+k)}(\vartheta,n)=F_{\text{min}}^{\mathcal{T}_{B}^{\alpha}|(\pm j),(\mp j+k)}(\vartheta,n) & \ \quad\forall i,j,k,\\
F_{\text{min}}^{\mathcal{T}_{B}^{\alpha}(\mp1),(\pm j)}(\vartheta,n)=e{}^{\mp i\alpha(j-1)/n}F_{\text{min}}^{\mathcal{T}_{B}^{\alpha}|(\pm1),(\mp1)}(2\pi i(j-1)-\vartheta,n) & \ \quad\forall j\neq1,
\end{split}
\label{eq:FMinAxiomsSemiLocalU1-1}
\end{equation}
and finally we get 
\begin{equation}
F_{\text{min}}^{\mathcal{T}_{B}^{\alpha}|(\pm j),(\mp k)}(\vartheta,n)=\begin{cases}
e^{\pm i\alpha(k-j)/n}F_{\text{min}}^{\mathcal{T}_{B}^{\alpha}|(\mp1),(\pm1)}(2\pi i(k-j)-\vartheta,n) & \text{ if }k>j,\\
e^{\mp i\alpha(j-k)/n}F_{\text{min}}^{\mathcal{T}_{B}^{\alpha}|(\pm1),(\mp1)}(2\pi i(j-k)+\vartheta,n) & \text{otherwise.}
\end{cases}\label{eq:FDMinBasicPropBoson}
\end{equation}
Akin to the previous case, the only independent quantity is $F_{\text{min}}^{\mathcal{T}_{B}^{\alpha}|(\pm1),(\mp1)}(\vartheta,n)$.
We exploit Eq. (\ref{eq:FMinAxiomsSemiLocalU1-1}) to write 
\begin{equation}
F_{\text{min}}^{\mathcal{T}_{B}^{\alpha}|(\pm1),(\mp1)}(\vartheta,n)=F_{\text{min}}^{\mathcal{T}_{B}^{\alpha}|(\mp1),(\pm1)}(-\vartheta,n)=e^{\pm i\alpha}F_{\text{min}}^{\mathcal{T}_{B}^{\alpha}|(\mp1),(\pm1)}(-\vartheta+2\pi in,n)\,.
\end{equation}
The solution of $F_{\text{min}}^{\mathcal{T}_{B}^{\alpha}|(\pm1),(\mp1)}$
can be obtained by introducing $f_{(\pm1),(\mp1)}^{\mathcal{T}_{B}^{\alpha}}(\vartheta)$
as 
\begin{equation}
F_{\text{min}}^{\mathcal{T}_{B}^{\alpha}|(\pm1),(\mp1)}(\vartheta,n)=f_{(\pm1),(\mp1)}^{\mathcal{T}_{B}^{\alpha}}(\vartheta/n)\,,
\end{equation}
that satisfies 
\begin{equation}
f_{(\pm1),(\mp1)}^{\mathcal{T}_{B}^{\alpha}}(\vartheta)=f_{(\pm1),(\mp1)}^{\mathcal{T}_{B}^{\alpha}}(-\vartheta)=e^{\pm i\alpha}f_{(\pm1),(\mp1)}^{\mathcal{T}_{B}^{\alpha}}(-\vartheta+2\pi i)\,.\label{f11axiomSemiLocalU1Boson}
\end{equation}
To obtain $f_{(\pm1),(\mp1)}^{\mathcal{T}_{B}^{\alpha}}$ it is instructive
to study first the conventional $U(1)$ twist field $\mathcal{V}_{B}^{\alpha}$.
The FFs of this field were computed in \cite{DoyonBosonU1VEV}, and
the two-particle FFs of this field read as 
\begin{equation}
F_{\text{2}}^{\mathcal{V}_{B}^{\alpha}|(\pm1),(\mp1)}(\vartheta,\alpha)=-\sin\frac{\alpha}{2}\frac{e^{\pm\frac{\vartheta(\alpha-\pi)}{2\pi}}}{\cosh\frac{\vartheta}{2}}\,,\label{F2U1B}
\end{equation}
where the range for $\alpha$ is now $[0,2\pi]$. From Eq. \eqref{F2U1B},
the minimal FF of the standard $U(1)$ field as well as $f_{(\pm1),(\mp1)}^{\mathcal{V}_{B}^{\alpha}}$
is inferred to be 
\begin{equation}
f_{(\pm1),(\mp1)}^{\mathcal{V}_{B}^{\alpha}}=-e^{\pm\frac{\vartheta(\alpha-\pi)}{2\pi}}\cosh\frac{\vartheta}{2}
\end{equation}
from which 
\begin{equation}
F_{\text{min}}^{\mathcal{T}_{B}^{\alpha}|(\pm1),(\mp1)}(\vartheta,n,\alpha)=-e^{\pm\frac{\vartheta(\alpha-\pi)}{2\pi n}}\cosh\frac{\vartheta}{2n}\,.
\end{equation}
It is straightforward to check that the ansatz (\ref{f11axiomSemiLocalU1Boson})
satisfies Eq. (\ref{eq:FFAxiom1SemiLocalGeneralBoson}) and Eq. (\ref{eq:FFAxiom2SemiLocalGeneralBoson}).

Similarly to the case of the Dirac theory, we can again write the
2-particle form factor as a product 
\begin{equation}
F_{2}^{\mathcal{T}_{B}^{\alpha}|(\pm j),(\mp k)}(\vartheta,n,\alpha)=P_{B}^{(\pm j),(\mp k)}(\vartheta,n,\alpha)F_{\text{min}}^{\mathcal{T}_{B}^{\alpha}|(\pm j),(\mp k)}(\vartheta,n,\alpha)\,,\label{eq:F2TU1ProductBoson}
\end{equation}
where $P$ satisfies 
\begin{equation}
P_{B}^{(\pm j),(\mp k)}(\vartheta,n,\alpha)=\begin{cases}
P_{B}^{(\mp1j),(\pm1)}(2\pi i(k-j)-\vartheta,n,\alpha) & \text{ if }k>j,\\
P_{B}^{(\pm1j),(\mp1)}(2\pi i(j-k)+\vartheta,n,\alpha) & \text{otherwise},
\end{cases}\label{eq:PBasicPropBoson}
\end{equation}
with 
\begin{equation}
P_{B}^{(+1),(-1)}(\vartheta,n,\alpha)=P_{B}^{(-1),(+1)}(-\vartheta,n,\alpha)\label{eq:ParitypropPBoson}
\end{equation}
together with the important property 
\begin{equation}
\begin{split}P_{B}^{(+1),(-1)}(2\pi ni+\vartheta,n,\alpha)=P_{B}^{(-1),(+1)}(-\vartheta,n,\alpha),\\
P_{B}^{(-1),(+1)}(2\pi ni+\vartheta,n,\alpha)=P_{B}^{(+1),(-1)}(-\vartheta,n,\alpha).
\end{split}
\label{eq:CyclicProeprtyPBoson}
\end{equation}
Combining Eq. \eqref{eq:FFAxiom3SemiLocalGeneralBoson} and the minimal
form factor, we have different expressions for the poles compared
to the Dirac case, namely 
\begin{equation}
\begin{split}-i\underset{x=0}{{\rm Res}}\,P_{B}^{(\pm1),(\mp1)}(x+i\pi,n,\alpha)\cosh\frac{i\pi}{2n}e^{\pm i(\alpha-\pi)/(2n)}= & 1\\
-i\underset{x=0}{{\rm Res}}\,P_{B}^{(\pm1),(\mp1)}(x-i\pi,n,\alpha)\cosh\frac{i\pi}{2n}e^{\mp i(\alpha-\pi)/(2n)}e^{\pm i\alpha/(n)}= & -e^{\pm i\alpha/(n)}\,.
\end{split}
\label{eq:Axiom3PBoson}
\end{equation}
We can further factorise $P_{B}^{(+1),(-1)}$ and $P_{B}^{(-1),(+1)}$
writing them as 
\begin{equation}
P_{B}^{(\pm1),(\mp1)}(\vartheta,n,\alpha)=p_{B}^{\pm}(\vartheta,n,\alpha)P(\vartheta,n,0)\,.
\end{equation}
$P(\vartheta,n,0)$ is chosen to be the usual one 
\begin{equation}
P(\vartheta,n,0)=\frac{\sin\frac{\pi}{n}}{2n\sinh\frac{i\pi-\vartheta}{2n}\sinh\frac{i\pi+\text{\ensuremath{\vartheta}}}{2n}\sinh\frac{i\pi}{2n}},
\end{equation}
but we must be more careful with the function $p_{B}^{\pm}(\vartheta,n,\vartheta)$.
In fact, to be compatible with \eqref{eq:FFAxiom3SemiLocalGeneralBoson}, it
must be $2\pi n$ periodic functions along the imaginary axis, whose
values at $i\pi$ and $-i\pi$ are $e^{\mp i(\alpha-\pi)/(2n)}$ and
$e^{\pm i(\alpha-\pi)/(2n)}$ respectively. 
The most natural choice for such a function is 
\begin{equation}
p_{B}^{\pm}(\vartheta,n,\vartheta)=\cos\frac{\alpha-\pi}{2n}\mp\sin\frac{\alpha-\pi}{2n}\frac{\sinh\frac{\vartheta}{n}}{\sin\frac{\pi}{n}}\,.\label{ppmBoson}
\end{equation}
With this choice, we re-obtain the standard $U(1)$ field, when $n\rightarrow1$.
We also have to recover the conventional branch-point twist field for $\alpha=0$, but unfortunately it is not the case. The latter requirement
can, nevertheless, be only satisfied if instead of $\sin\frac{\alpha-\pi}{2n}\frac{\sinh\frac{\vartheta}{n}}{\sin\frac{\pi}{n}}$
in Eq. \eqref{ppmBoson}, we use $\sin\frac{\alpha-\pi}{2n}\sin\frac{\pi}{n}\sinh^{2}\frac{\vartheta}{2n}/(\sinh\frac{\vartheta}{n}\sin^{2}\frac{\pi}{2n})$.
Note that, interestingly the function that multiplies $\sin\frac{\alpha-\pi}{2n}$
is exactly the inverse of the corresponding function for the Dirac
theory, which multiplies $\sin\frac{\alpha}{2n}$ in $p^{\pm}$. Eventually,
we arrive at our final expression for $P_{B}^{(\pm1),(\mp1)}$, namely
\begin{equation}
P_{B}^{(\pm1),(\mp1)}(\vartheta,n,\vartheta)=\frac{\sin\frac{\pi}{n}}{2n\sinh\frac{i\pi-\vartheta}{2n}\sinh\frac{i\pi+\text{\ensuremath{\vartheta}}}{2n}\cosh\frac{i\pi}{2n}}\left(\cos\frac{\alpha-\pi}{2n}\mp\sin\frac{\alpha-\pi}{2n}\frac{\sin\frac{\pi}{n}\sinh^{2}\frac{\vartheta}{2n}}{\sinh\frac{\vartheta}{n}\sin^{2}\frac{\pi}{2n}}\right)\,.\label{eq:PSolutionBoson}
\end{equation}
Using the formula \eqref{eq:PSolutionBoson} to construct the 2-particle
form factors via \eqref{eq:F2TU1ProductBoson} and \eqref{eq:PBasicPropBoson}
all the axioms \eqref{eq:FFAxiom1SemiLocalGeneralBoson}, \eqref{eq:FFAxiom2SemiLocalGeneralBoson},
\eqref{eq:FFAxiom3SemiLocalGeneralBoson} are satisfied and the $\alpha\rightarrow0$
and $n\rightarrow1$ limits yield the desired results. i.e., the conventional
branch-point and the standard $U(1)$ twist field FFs. After algebraic simplifications,
the full 2-particle FF can be therefore written as 
\begin{equation}
F_{2}^{\mathcal{T}_{B}^{\alpha}|(\pm1),(\mp1)}(\vartheta,n,\alpha)=\frac{\langle\mathcal{T}_{B,n}^{\alpha}\rangle\sin\frac{\pi}{n}}{2n\sinh\frac{i\pi-\vartheta}{2n}\sinh\frac{i\pi+\text{\ensuremath{\vartheta}}}{2n}}\left(\frac{\cos\frac{\alpha-\pi}{2n}\cosh\frac{\vartheta}{2n}}{\cos\frac{\pi}{2n}}\mp\frac{\sin\frac{\alpha-\pi}{2n}\sinh\frac{\vartheta}{2n}}{\sin\frac{\pi}{2n}}\right)e^{\pm\frac{\vartheta(\alpha-\pi)}{2\pi n}},\label{eq:F2BosonSolution11}
\end{equation}
where the range of $\alpha$ is $[0,2\pi]$. We can conform our results
to the Dirac case and shift the range of $\alpha$ to $[-\pi,\pi]$
by redefining our expression as
\begin{equation}
\begin{split}F_{2}^{\mathcal{T}_{B}^{\alpha}|(\pm1),(\mp1)}(\vartheta,n,\alpha)= & \frac{\langle\mathcal{T}_{B,n}^{\alpha}\rangle\sin\frac{\pi}{n}}{2n\sinh\frac{i\pi-\vartheta}{2n}\sinh\frac{i\pi+\text{\ensuremath{\vartheta}}}{2n}}\begin{cases}
\left(\frac{\cos\frac{\alpha-\pi}{2n}\cosh\frac{\vartheta}{2n}}{\cos\frac{\pi}{2n}}\mp\frac{\sin\frac{\alpha-\pi}{2n}\sinh\frac{\vartheta}{2n}}{\sin\frac{\pi}{2n}}\right)e^{\pm\frac{\vartheta(\alpha-\pi)}{2\pi n}} & \text{if }\alpha\geq0\\
\left(\frac{\cos\frac{\alpha+\pi}{2n}\cosh\frac{\vartheta}{2n}}{\cos\frac{\pi}{2n}}\mp\frac{\sin\frac{\alpha+\pi}{2n}\sinh\frac{\vartheta}{2n}}{\sin\frac{\pi}{2n}}\right)e^{\pm\frac{\vartheta(\alpha+\pi)}{2\pi n}} & \text{if }\alpha<0.
\end{cases}\end{split}
\end{equation}
Finally, using Eq. \eqref{eq:FDMinBasicProp}, we get 
\begin{equation}
F_{2}^{\mathcal{T}_{B}^{\alpha}|(\pm j),(\mp k)}(\vartheta,n)=\begin{cases}
e^{\pm i\alpha(k-j)/n}F_{2}^{\mathcal{T}_{B}^{\alpha}|(\mp1),(\pm1)}(2\pi i(k-j)-\vartheta,n) & \text{ if }k>j,\\
e^{\mp i\alpha(j-k)/n}F_{2}^{\mathcal{T}_{B}^{\alpha}|(\pm1),(\mp1)}(2\pi i(j-k)+\vartheta,n) & \text{otherwise}\,,
\end{cases}\label{eq:FB2Full}
\end{equation}

\subsection{$\Delta$-theorem test and higher-particle FFs}

The validity of the solution can be checked by the $\Delta$-theorem
\begin{equation}
-\frac{n}{32\pi^{2}m^{2}\langle\mathcal{T}_{D}^{\alpha}\rangle}\int\mathrm{d}\vartheta\frac{F_{B,2}^{\Theta|1,1}\left(\vartheta\right)\left(F_{2}^{\mathcal{T}_{B}^{\alpha}|(\pm1),(\mp1)}(\vartheta,n)^{*}+F_{2}^{\mathcal{T}_{B}^{\alpha}|(\mp1),(\pm1)}(\vartheta,n)^{*}\right)}{\cosh^{2}\left(\vartheta/2\right)}\,.
\end{equation}
The chiral conformal dimensions obtained from the appropriate integrals
of the 2-particle FFs gives the exact UV result \cite{mdc-20b}
\begin{equation}
\Delta_{n}^{\mathcal{T}_{B}^{\alpha}}=\frac{2}{24}\left(n-n^{-1}\right)+\frac{\left(\left|\alpha\right|/(2\pi)\right)\left(1-\left|\alpha\right|/(2\pi)\right)}{2n}\,.
\end{equation}

The FFs of higher particle numbers are obtained by using Wick's theorem. Assuming the ordering of the replica index $\mu_{1}\geq\mu_{2}\geq...\geq\mu_{2k}$,
one has 
\begin{equation}
F_{2k}^{\mathcal{T}_{B}^{\alpha}|\underline{(\beta\mu)}}(\underline{\vartheta})=\langle\mathcal{T}_{B}^{\alpha}\rangle\sum_{\text{all pairings}}\prod_{\{l.m\}}\frac{F_{2}^{\mathcal{T}_{B}^{\alpha}|(\beta_{l}\mu_{l}),(\beta_{m}\mu_{m})}(\vartheta_{l}-\vartheta_{m},n)}{\langle\mathcal{T}_{B}^{\alpha}\rangle}\,,\label{eq:Wick}
\end{equation}
where $\{l,m\}$ runs over the possible pairs and $F_{2}^{\mathcal{T}_{B}^{\alpha}|(\beta_{l}\mu_{l}),(\beta_{m}\mu_{m})}$
is identically zero if $\beta_{l}=\beta_{m}=\pm1$ due to the neutrality
condition of the composite $U(1)$ branch-point twist field. 

Similarly to the Dirac theory, the FFs of the field $\tilde{\mathcal{T}}_{B}^{\alpha}$
can be again easily obtained from that of $\mathcal{T}_{B}^{\alpha}$.
As long as the range of $\alpha$ is set to $[-\pi,\pi]$, the 2-particle
FF can be written as
\begin{equation}
F_{2}^{\tilde{\mathcal{T}}_{B}^{\alpha}|(\pm j),(\mp k)}(\vartheta,n,\alpha)=F_{2}^{\mathcal{T}_{B}^{-\alpha}|(\pm n-j),(\mp n-k)}(\vartheta,n,-\alpha)\,.
\end{equation}

\section{Form factors of the composite U(1) branch-point twist field via diagonalisation
in replica space \label{secAlternative}}

In this section, we provide an
alternative derivation of the two-particle FF of the modified twist
fields, based on the diagonalisation in the space of replicas \cite{CFH}.
This technique has been already employed in \cite{cdds-18b} for the
computation of the FF of the standard branch-point twist fields. It
works well for free theories when the $S$-matrix does not depend
explicitly on the rapidities and it is more closely related to the
approach in Ref. \cite{mdc-20b}. We will briefly summarise this formalism,
discussing the case in which a non-vanishing flux is inserted.

 Let us consider the creation operator $A_{j,\pm}^{\dagger}(\vartheta)$
of a particle/antiparticle in the $j$-th replica. The cyclic symmetry
of replicas can be diagonalised moving to a ``replica Fourier space''
\begin{equation}
\hat{A}_{k,\pm}^{\dagger}(\vartheta)=\frac{1}{\sqrt{n}}\sum_{j=0}^{n-1}e^{\mp i\frac{2\pi kj}{n}}A_{j,\pm}^{\dagger}(\vartheta),\label{dFZ}
\end{equation}
where we identified $j\sim j+n$ and $k\sim k+n$. In order to account
for the correct (anti)commutation relations (and hence $S$-matrix)
of bosons and fermions, the index $k$ should be either integer or
half-integer according to the following rules 
\begin{itemize}
\item for bosons ($S(\vartheta)=1$) $k=0,\dots,n-1$; 
\item for fermions ($S(\vartheta)=-1$) $k=-\frac{n-1}{2},\dots,\frac{n-1}{2}$. 
\end{itemize}
Actually, different prescriptions can be found in the literature for
the diagonalisation in the replica space \cite{ccd-08,CFH,mdc-20b,ch-rev,ch-05}.
These prescriptions are all related to each other by unitary transformations
and in the end the diagonal modes simply get a phase. Since here we
are only interested in the absolute value squared of the form-factors,
this issue is irrelevant and any prescription would give the same
result. The advantage of this approach is that the modified twist
field $\mathcal{T}_{n}^{\alpha}$ does not mix different Fourier modes
and, consequently, its non-vanishing 2-particle form factors are just
\begin{equation}
\mathcal{F}_{k,\pm}^{\alpha}(\vartheta)=\langle0|\mathcal{T}_{n}^{\alpha}(0)\hat{A}_{k,\pm}^{\dagger}(\vartheta)\hat{A}_{k,\mp}^{\dagger}(0)|0\rangle.\label{FDD}
\end{equation}
These FF have to satisfy the following bootstrap equations 
\begin{equation}
\mathcal{F}_{k,\pm}^{\alpha}(\vartheta+2\pi i)=e^{\pm i\frac{\alpha}{n}\pm\frac{i2\pi k}{n}}\mathcal{F}_{k,\mp}^{\alpha}(-\vartheta),
\end{equation}
\begin{equation}
\underset{\vartheta=i\pi}{{\rm Res}}\,\mathcal{F}_{k,\pm}^{\alpha}(\vartheta)=i(1-e^{\pm i\frac{\alpha}{n}\pm\frac{i2\pi k}{n}})\langle\mathcal{T}_{n}^{\alpha}\rangle,
\end{equation}
\begin{equation}
\mathcal{F}_{k,\pm}^{\alpha}(-\vartheta)=S(\vartheta)\mathcal{F}_{k,\mp}^{\alpha}(\vartheta),
\end{equation}
which obviously differ from the equations given in \cite{cdds-18b} for the conventional twist field. The only difference is the 
 additional phase shift $e^{\pm i\alpha/n}$, as a consequence of the fact that  $\mathcal{T}_{n}^{\alpha}$
introduces a phase $e^{i\alpha/n}$ between each couple of consecutive replicas. 
(In principle, other choices for the phase of the single field are possible with the constraint that the total
phase is $e^{i\alpha}$; however, these different choices lead to different Fourier modes \eqref{dFZ} for $\mathcal{T}_{n}^{\alpha}$.) 
For simplicity we will focus on $\mathcal{F}_{k,+}^{\alpha}$ for
$\alpha>0$, since the other cases are obtained by symmetry. Given
the form factors \eqref{FDD} in Fourier space, the physical ones
are obtained as sum over the Fourier modes. In particular one has
\begin{multline}
  \langle0|\mathcal{T}_{n}^{\alpha}(0)A_{j,+}^{\dagger}(\vartheta)A_{j,-}^{\dagger}(0)|0\rangle=\\
  \langle0|\mathcal{T}_{n}^{\alpha}(0)\left(\frac{1}{\sqrt{n}}\sum_{k}\hat{A}_{k,+}^{\dagger}(\vartheta)e^{+i\frac{2\pi kj}{n}}\right)\left(\frac{1}{\sqrt{n}}\sum_{k'}\hat{A}_{k',-}^{\dagger}(\vartheta)e^{-i\frac{2\pi k'j}{n}}\right)|0\rangle=\frac{1}{n}\sum_{k}\mathcal{F}_{k,+}^{\alpha}(\vartheta).\label{ffff}
\end{multline}

\subsection{Complex bosons}

A solution of the bootstrap equations for the complex boson is 
\begin{equation}
\mathcal{F}_{k,+}^{\alpha}(\vartheta)=-\langle\mathcal{T}_{n}^{\alpha}\rangle\sin\left(\frac{\alpha}{2n}+\frac{\pi k}{n}\right)\frac{e^{\alpha\vartheta/2\pi n+k\vartheta/n-\vartheta/2}}{\cosh\vartheta/2}.
\end{equation}
The sum over the modes \eqref{ffff} is easily done using 
\begin{equation}
\sum_{k=0}^{n-1}e^{i\gamma k}=\frac{e^{i\gamma n}-1}{e^{i\gamma}-1}=e^{\frac{i\gamma}{2}(n-1)}\frac{\sin\frac{n\gamma}{2}}{\sin\frac{\gamma}{2}},
\end{equation}
to get 
\begin{equation}
\langle0|\mathcal{T}_{n}^{\alpha}(0)A_{1,+}^{\dagger}(\vartheta)A_{1,-}^{\dagger}(0)|0\rangle=-i\langle\mathcal{T}_{n}^{\alpha}\rangle\frac{e^{\alpha\vartheta/2\pi n-\vartheta/2n}}{2n\sinh\frac{\vartheta+i\pi}{2n}\sinh\frac{\vartheta-i\pi}{2n}}\left(e^{i\alpha/2n-i\pi/2n}\sinh\frac{\vartheta-i\pi}{2n}-(\text{c.c.})\right),
\end{equation}
which corresponds to the result \eqref{eq:F2BosonSolution11} by standard methods.
This is the physical solution only for $\alpha>0$,
while the one for $\alpha<0$ can be obtained by symmetry. One can
derive the same expression starting from the ansatz 
\begin{equation}
\langle0|\mathcal{T}_{n}^{\alpha}(0)A_{1,+}^{\dagger}(\vartheta)A_{1,-}^{\dagger}(0)|0\rangle=\langle\mathcal{T}_{n}^{\alpha}\rangle\frac{e^{\alpha\vartheta/2\pi n-\vartheta/2n}}{2n\sinh\frac{\vartheta+i\pi}{2n}\sinh\frac{\vartheta-i\pi}{2n}}\left(C_{0}e^{-\vartheta/2n}+C_{1}e^{\vartheta/2n}\right),
\end{equation}
compatible with the monodromy equations, and choosing $C_{0},C_{1}$
such that the poles are 
\begin{equation}
\underset{\vartheta=i\pi}{{\rm Res}}\,\langle0|\mathcal{T}_{n}^{\alpha}(0)A_{1,+}^{\dagger}(\vartheta)A_{1,-}^{\dagger}(0)|0\rangle =i\langle\mathcal{T}_{n}^{\alpha}\rangle,
\end{equation}
\begin{equation}
\underset{\vartheta=2in\pi-i\pi}{{\rm Res}}\langle0|\mathcal{T}_{n}^{\alpha}(0)A_{1,+}^{\dagger}(\vartheta)A_{1,-}^{\dagger}(0)|0\rangle =-ie^{i\alpha}\langle\mathcal{T}_{n}^{\alpha}\rangle.
\end{equation}
This result is valid also when analytically continued to $n\rightarrow1$,
and provides the FF of $\mathcal{V}_{\alpha}$ (nonzero if $\alpha\neq0$).
Indeed, in that limit, the poles at $\vartheta=i\pi,2in\pi-i\pi$
collapse together with an additional zero so that a single pole is
left at $\vartheta=i\pi$, as a consequence of 
\begin{equation}
\pm\frac{\sinh\frac{\vartheta\pm i\pi}{2}}{\sinh\frac{\vartheta+i\pi}{2}\sinh\frac{\vartheta-i\pi}{2}}=\pm\frac{1}{\sinh\frac{\vartheta-i\pi}{2}}.
\end{equation}
This leftover pole is exactly the one for the single replica model, namely 
\begin{equation}
\underset{\vartheta=i\pi}{{\rm Res}}\,\langle0|\mathcal{V}^{\alpha}(0)A_{1,+}^{\dagger}(\vartheta)A_{1,-}^{\dagger}(0)|0\rangle=i(1-e^{i\alpha})\langle\mathcal{V}^{\alpha}\rangle.
\end{equation}

\subsection{Dirac fermions}

Similar calculations can be done for free fermions, which lead to 
\begin{equation}
\mathcal{F}_{k,+}^{\alpha}(\vartheta)=i\langle\mathcal{T}_{n}^{\alpha}\rangle\sin\left(\frac{\alpha}{2n}+\frac{\pi k}{n}\right)\frac{e^{\alpha\vartheta/2\pi n+k\vartheta/n}}{\cosh\frac{\vartheta}{2}}.
\end{equation}
The sum over the modes is performed using 
\begin{equation}
\sum_{k=-\frac{n-1}{2}}^{\frac{n-1}{2}}e^{i\gamma k}=\frac{e^{i\gamma n/2}-e^{-i\gamma n/2}}{e^{i\gamma n/2}-e^{-i\gamma n/2}}=\frac{\sin\frac{n\gamma}{2}}{\sin\frac{\gamma}{2}},
\end{equation}
to obtain 
\begin{equation}
\begin{split}
 & \langle0|\mathcal{T}_{n}^{\alpha}(0)A_{1,+}^{\dagger}(\vartheta)A_{1,-}^{\dagger}(0)|0\rangle=\\
 & -i\langle\mathcal{T}_{n}^{\alpha}\rangle\frac{e^{\alpha\vartheta/2\pi n}}{2n\sinh\frac{i\pi+\vartheta}{2n}\sinh\frac{i\pi-\vartheta}{2n}}\left(e^{i\alpha/2n}\sinh\frac{\vartheta-i\pi}{2n}+e^{-i\alpha/2n}\sinh\frac{\vartheta+i\pi}{2n}\right),
\end{split}
\end{equation}
and this provides an alternative derivation of the FF in Eq. \eqref{eq:F2DiracSolution11}.
 
 As for the case of complex boson, this solution can be also obtained making the ansatz
\begin{equation}
\langle0|\mathcal{T}_{n}^{\alpha}(0)A_{1,+}^{\dagger}(\vartheta)A_{1,-}^{\dagger}(0)|0\rangle=\langle\mathcal{T}_{n}^{\alpha}\rangle\frac{e^{\alpha\vartheta/2\pi n}}{2n\sinh\frac{\vartheta+i\pi}{2n}\sinh\frac{\vartheta-i\pi}{2n}}\left(C_{0}e^{-\vartheta/2n}+C_{1}e^{\vartheta/2n}\right),
\end{equation}
and fixing $C_{0},C_{1}$ compatibly with the kinematical residues.

\section{U(1) charged moments in free massive QFTs\label{secChargedMoments}}

\label{sec:gen}

In this section, we first present some basic and elementary facts
about the $U(1)$ charged moments $Z_{n}(\alpha)$ for free theories
with $U(1)$ symmetry. Exploiting the QFT scaling form, some of our
results are valid for arbitrary massive QFTs with $U(1)$ symmetry
with free bosonic/fermionic type UV limiting CFT as well. We restrict
our analysis to a subsystem composed of a single interval and the
full quantum system is prepared in its ground state. In this setting,
the charged moments and later on entropies as well can be calculated
from the two-point functions of the $U(1)$ composite twist fields,
which we explicitly calculate in the following. Specifying the subsystem
as an interval $A=[u,v]$ (with $\ell=v-u$), the charged moments
are written as 
\begin{eqnarray}
Z_{n}(\alpha) & = & \text{Tr}\left(\rho_{A}^{n}e^{i\alpha\hat{Q}_{A}}\right)=\zeta_{n}^{\alpha}\varepsilon^{2d_{n}^{\alpha}}\langle\mathcal{T}_{n}^{\alpha}(u,0)\tilde{\mathcal{T}}_{n}^{\alpha}(v,0)\rangle\,,\label{eq:Zn0TwoPt}
\end{eqnarray}
where $\varepsilon$ is the UV regulator, $\zeta_{n}^{\alpha}$ is
a normalisation constant for the charged moments and $d_{n}^{\alpha}$
is the scaling dimension of the composite twist field 
\begin{equation}
d_{n}^{\alpha}=
2\Delta_{n}^{\mathcal{T}^{\alpha}}=\begin{cases}
\frac{1}{12}\left(n-n^{-1}\right)+\frac{\alpha^{2}}{\left(2\pi\right)^{2}n}, & \text{Dirac},\\
\frac{1}{6}\left(n-n^{-1}\right)-\frac{\alpha^{2}}{(2\pi)^{2}n}+\frac{|\alpha|}{2\pi n}, & \text{complex Boson}.
\end{cases}
\end{equation}
To keep track of the leading $\ell$-dependence as well, we rewrite
the charged moments as 
\begin{equation}
Z_{n}(\alpha)=\zeta_{n}^{\alpha}\varepsilon^{2d_{n}^{\alpha}}\langle\mathcal{T}_{n}^{\alpha}(u,0)\tilde{\mathcal{T}}_{n}^{\alpha}(v,0)\rangle\equiv\zeta_{n}^{\alpha}(m\varepsilon)^{2d_{n}^{\alpha}}[(m^{-2d_{n}^{\alpha}}\langle\mathcal{T}_{n}^{\alpha}\rangle^{2})]H_{\alpha,n}(m\ell)\,.\label{znU1}
\end{equation}
Using the standard two-particle approximation, the scaling function
of the two-point correlation, $H_{\alpha,n}(m\ell)$, for generic
$n$ can be written as 
\begin{equation}
H_{\alpha,n}^{{\rm 2pt}}(m\ell)=1+\frac{n}{4\pi^{2}}\int_{-\infty}^{\infty}\mathrm{d}\vartheta f_{D/B}^{\alpha}(\vartheta,n)K_{0}(2m\ell\cosh(\vartheta/2))\,,\label{H2pt}
\end{equation}
due to the expansion of the 2-point function 
\begin{equation}
\begin{split}\langle\mathcal{T}_{n}^{\alpha}(\ell,0)\tilde{\mathcal{T}}_{n}^{\alpha}(0,0)\rangle\approx & \langle\mathcal{T}_{n}^{\alpha}\rangle^{2}+\sum_{j,k=1}^{n}\int_{-\infty}^{\infty}\frac{\mathrm{d}\vartheta_{1}\mathrm{d}\vartheta_{2}}{(2\pi)^{2}2!}|F_{2}^{\mathcal{T}^{\alpha}|(+j),(-k)}(\vartheta_{12},n)|^{2}e^{-rm\left(\cosh\vartheta_{1}+\cosh\vartheta_{2}\right)}\\
 & \qquad+\sum_{j,k=1}^{n}\int_{-\infty}^{\infty}\frac{\mathrm{d}\vartheta_{1}\mathrm{d}\vartheta_{2}}{(2\pi)^{2}2!}|F_{2}^{\mathcal{T}^{\alpha}|(-j),(+k)}(\vartheta_{12},n)|^{2}e^{-rm\left(\cosh\vartheta_{1}+\cosh\vartheta_{2}\right)}\\
= & \langle\mathcal{T}_{n}^{\alpha}\rangle^{2}\left(1+\frac{n}{4\pi^{2}}\int_{-\infty}^{\infty}\mathrm{d}\vartheta f_{D/B}^{\alpha}(\vartheta,n)K_{0}\left(2m\ell\cosh\left(\vartheta/2\right)\right)\right)\,.
\end{split}
\label{2ptf}
\end{equation}
In the above formula $f_{D/B}^{\alpha}(\vartheta,n)$ is implicitly
defined as 
\begin{equation}
\begin{split}\langle\mathcal{T}_{\kappa,n}^{\alpha}\rangle^{2}f_{\kappa}^{\alpha}(\vartheta,n)= & \sum_{j=1}^{n}|F_{2}^{\mathcal{T}_{\kappa}^{\alpha}|(+1),(-j)}(\vartheta,n)|^{2}+|F_{2}^{\mathcal{T}_{\kappa}^{\alpha}|(-1),(+j)}(\vartheta,n)|^{2}\\
= & \sum_{j=0}^{n-1}|F_{2}^{\mathcal{T}_{\kappa}^{\alpha}|(-1),(+1)}(2\pi ij-\vartheta,n)|^{2}+\sum_{j=0}^{n-1}|F_{2}^{\mathcal{T}_{\kappa}^{\alpha}|(+1),(-1)}(2\pi ij-\vartheta,n)|^{2}\,.
\end{split}
\label{eq:f(theta,n)}
\end{equation}
To ease our notations, we introduced the index $\kappa$ which is
either $D$ or $B$ referring to the Dirac fermion and complex boson respectively.
The calculation of the von-Neumann entropy (and of the corresponding
charged moments) requires the correct analytic continuation $\tilde{f}_{\kappa}^{\alpha}(\vartheta,n)$
of the function $f_{\kappa}^{\alpha}(\vartheta,n)$, in such a way
to justify the treatment of $n$ as a continuous variable. While for any integer
$n\geq2$, $f_{\kappa}^{\alpha}(\vartheta,n)=\tilde{f}_{\kappa}^{\alpha}(\vartheta,n)$,
this equality is no longer true at $n=1$: $\tilde{f}_{\kappa}^{\alpha}(\vartheta,1)$
equals $f_{\kappa}^{\alpha}(\vartheta,1)$ for $\kappa=D,B$, any
$\alpha$ and $\vartheta\neq0$. In other words $\tilde{f}_{\kappa}^{\alpha}(\vartheta,1)$
is not a continuous function in $\vartheta$. An important consequence
of this fact is that the derivative of $f_{\kappa}^{\alpha}(\vartheta,n)$
wrt. $n$ contains a $\delta(\vartheta)$ contribution. See appendix
\ref{sec:AppendixA-AnalContForfDB} for more details.

Eq. \eqref{H2pt} with \eqref{eq:f(theta,n)} provides an explicit
final result for the Rényi entropies with integer $n>1$. In fact for the leading spatial dependence, we can
use the small $\vartheta$ behaviour of the $f_{\kappa}^{\alpha}(\vartheta,n)$
derived also in appendix \ref{sec:AppendixA-AnalContForfDB}. 
Plugging the expansions in Eqs. \eqref{AnConFTildenDirac} and \eqref{AnConFTildenBoson} into the integral in Eq. \eqref{2ptf} and using \cite{Z2IsingShg} 
\begin{equation}
\frac{1}{2\pi^{2}}\int_{-\infty}^{\infty}\text{d}\vartheta K_{0}\left(2m\ell\cosh\frac{\vartheta}{2}\right)\left(\frac{\vartheta}{2}\right)^{2n}=\frac{\Gamma\left(n+\frac{1}{2}\right)}{2\pi^{3/2}}\frac{e^{-2m\ell}}{(m\ell)^{n+1}}(1+O((m\ell)^{-1}),\label{eq:BesselTheta2n}
\end{equation}
we immediately have 
\begin{equation}
H_{\alpha,n}^{{\rm 2pt}}(m\ell)=1+\frac{n}{4\pi}\frac{e^{-2m\ell}}{m\ell}(1+O((m\ell)^{-1}),\label{H22}
\end{equation}
for $n>1$ and for both $\kappa=D,B$. Notably the leading order of
$H_{\alpha,n}^{{\rm 2pt}}(m\ell)$ is $\alpha$-independent.

As already mentioned for the limit $n\to1$ and for the derivative
of $f_{\kappa}^{\alpha}(\vartheta,n)$ wrt. $n$, we have to use the
proper the analytic continuation $\tilde{f}_{\kappa}^{\alpha}(\vartheta,1)$
of $f_{\kappa}^{\alpha}(\vartheta,n)$. Using the Taylor expansion
in $\vartheta$ of the analytic parts of $\tilde{f}_{\kappa}^{\alpha}(\vartheta,1)$
(that can be found in the appendix as Eqs. \eqref{fTildeDiracn1AnalyticPart}
and \eqref{fTildeBosonAnalyticpart}), the leading behaviour is determined
by the constant $2\sin^{2}\frac{\alpha}{2}$. Therefore we have the $\kappa$-independent
expression for $H_{\alpha,1}^{{\rm 2pt}}$ 
\begin{equation}
H_{\alpha,1}^{{\rm 2pt}}(m\ell)=1+2\sin^{2}\frac{\alpha}{2}\frac{1}{4\pi}\frac{e^{-2m\ell}}{m\ell}(1+O((m\ell)^{-1})\,,
\end{equation}
manifesting the discontinuous behaviour compared to Eq. \eqref{H22}. For
the derivative of $H_{\alpha,n}^{{\rm 2pt}}$, we have 
\begin{equation}
\begin{split}\frac{\partial}{\partial n}H_{\alpha,n}^{{\rm 2pt}}(m\ell)|_{n=1}= & \frac{1}{4}\text{\ensuremath{\cos\alpha}}K_{0}(2m\ell)+2\sin^{2}\frac{\alpha}{2}\frac{e^{-2m\ell}}{4\pi m\ell}(1+O((m\ell)^{-1})\\
= & \text{\ensuremath{\cos\alpha}}\frac{\sqrt{\pi}}{8}\frac{e^{-2m\ell}}{\sqrt{m\ell}}+2\sin^{2}\frac{\alpha}{2}\frac{e^{-2m\ell}}{4\pi m\ell}+O(e^{-2m\ell}/(m\ell)^{3/2}),
\end{split}
\end{equation}
again for both statistics $\kappa$. Here the first term comes from
the $\delta(\vartheta)$ in ${\displaystyle \lim_{n\rightarrow1}\frac{\partial}{\partial n}\tilde{f}_{\kappa}^{\alpha}(\vartheta,n)}$
and in the second line we also used that $K_{0}(z)\approx e^{-z}\sqrt{\frac{\pi}{2z}}$.

Summarising, the leading spatial dependence of charged moments for
$n\geq2$ can be written as 
\begin{equation}
Z_{n}(\alpha)=\zeta_{\kappa,n}^{\alpha}(m\varepsilon)^{2d_{\kappa,n}^{\alpha}}[(m^{-2d_{\kappa,n}^{\alpha}}\langle\mathcal{T}_{\kappa,n}^{\alpha}\rangle^{2})]\left(1+\frac{n}{4\pi}\frac{e^{-2m\ell}}{m\ell}(1+O((m\ell)^{-1})\right),\label{ZnU(1)}
\end{equation}
where $\kappa=D,B$, $\zeta_{\kappa,n}^{\alpha}$ is a normalisation
constant, $\varepsilon$ is a UV regulator and the VEV of the composite
twist fields is calculated in appendices \ref{sec:AppendixB-VEVDirac}
and \ref{sec:AppendixC-VEVBoson} for fermions and bosons, respectively.
For $n=1$, we have instead 
\begin{equation}
Z_{1}(\alpha)=\zeta_{\kappa,1}^{\alpha}(m\varepsilon)^{2d_{\kappa,1}^{\alpha}}[(m^{-2d_{\kappa,1}^{\alpha}}\langle\mathcal{T}_{\kappa,1}^{\alpha}\rangle^{2})]\left(1+2\sin^{2}\frac{\alpha}{2}\,\frac{1}{4\pi}\frac{e^{-2m\ell}}{m\ell}(1+O((m\ell)^{-1})\right)\,,\label{Z1U(1)}
\end{equation}
and for the derivative 
\begin{equation}
\begin{split}\frac{\partial}{\partial n}Z_{1}(\alpha)\Big|_{n=1}= & \left(\frac{1}{3}-\frac{\alpha^{2}}{2\pi^{2}}\right)(m\varepsilon)^{\frac{\alpha^{2}}{2\pi^{2}}}\ln(m\varepsilon)\zeta_{\kappa,1}^{\alpha}[(m^{-\frac{\alpha^{2}}{2\pi^{2}}}\langle\mathcal{T}_{\kappa,1}^{\alpha}\rangle^{2})]\left(1+2\sin^{2}\frac{\alpha}{2}\,\frac{e^{-2m\ell}}{4\pi m\ell}\right)\\
+ & \zeta_{\kappa,1}^{\alpha}(m\varepsilon)^{\frac{\alpha^{2}}{2\pi^{2}}}[(m^{-\frac{\alpha^{2}}{2\pi^{2}}}\langle\mathcal{T}_{\kappa,1}^{\alpha}\rangle^{2})]\left(\text{\ensuremath{\cos\alpha}}\frac{\sqrt{\pi}}{8}\frac{e^{-2m\ell}}{\sqrt{m\ell}}+2\sin^{2}\frac{\alpha}{2}\,\frac{e^{-2m\ell}}{4\pi m\ell}\right)\\
+ & \zeta_{\kappa,1}^{\alpha}(m\varepsilon)^{\frac{\alpha^{2}}{2\pi^{2}}}[(m^{-\frac{\alpha^{2}}{2\pi^{2}}}\langle\mathcal{T}_{\kappa,1}^{\alpha}\rangle^{2})]\left(1+2\sin^{2}\frac{\alpha}{2}\,\frac{e^{-2m\ell}}{4\pi m\ell}\right)\times\\
 & \left(\frac{\partial}{\partial n}\ln\zeta_{\kappa,n}^{\alpha}|_{n=1}+\frac{\partial}{\partial n}\ln[(m^{-2d_{\kappa,n}^{\alpha}}\langle\mathcal{T}_{\kappa,n}^{\alpha}\rangle^{2})]|_{n=1}\right)\,.
\end{split}
\label{DnZnU(1)}
\end{equation}
These formulas are the final results of this paper, which are derived
based solely on the form factor bootstrap. They perfectly match with
the results of Ref. \cite{mdc-20b} derived by completely different
means.

\section{U(1) symmetry resolved partition functions and entropies\label{sec:CalculatingEntropies} }

For free theories, the symmetry resolved partition functions and entropies
have been already calculated in Ref. \cite{mdc-20b} from the charged
moments that were equivalent to those in the previous section. It
is however useful to recall some steps of this derivation within the
notations of this paper, because the computation of the SR entropies
from charged moments is non-trivial.

Let us start recalling the definition of the symmetry resolved partition
functions (\ref{eq:CalU1}) in terms the charged moments (\ref{eq:Zn}):
\begin{equation}
\mathcal{Z}_{n}(q_{A})=\int_{-\pi}^{\pi}\frac{\mathrm{d}\alpha}{2\text{\ensuremath{\pi}}}Z_{n}(\alpha)e^{-i\alpha q_{A}}.
\end{equation}
To perform the Fourier transform of Eqs. \eqref{ZnU(1)}-\eqref{DnZnU(1)},
the knowledge of both $\langle\mathcal{T}_{\kappa,n}^{\alpha}\rangle$
and $\zeta_{\kappa,n}^{\alpha}$ is required as well. Although we
computed the VEV of the composite twist-fields (see appendices \ref{sec:AppendixB-VEVDirac}
and \ref{sec:AppendixC-VEVBoson}), which is a universal quantity
(once the UV normalization is fixed), there is no general recipe to
obtain $\zeta_{\kappa,n}^{\alpha}$ since it is non-universal and
its knowledge does not rely on QFT techniques.

Following Ref. \cite{mdc-20b}, we can write the logarithm of the
charged moments \eqref{ZnU(1)}-\eqref{DnZnU(1)} for both fermions/bosons
as 
\begin{equation}
\begin{split}\ln Z_{n}(\alpha)\simeq & \ln Z_{n}^{(0)}(\alpha)+\frac{ne^{-2m\ell}}{4\pi m\ell}+\dots,\\
\ln Z_{1}(\alpha)\simeq & \ln Z_{1}^{(0)}(\alpha)+\frac{e^{-2m\ell}}{2\pi m\ell}\sin^{2}\frac{\alpha}{2}+\dots,\\
\frac{d}{dn}\ln Z_{n}(\alpha)|_{n\rightarrow1^{+}}\simeq & \frac{d}{dn}\ln Z_{n}^{(0)}(\alpha)|_{n\rightarrow1^{+}}+\frac{1}{4}\cos\alpha K_{0}(2m\ell)+\dots,
\end{split}
\label{eq:LogZnU(1)Approx}
\end{equation}
where $\ln Z_{n}^{(0)}(\alpha)$ is 
\begin{align}
\ln Z_{n}^{(0)}(\alpha) & =\left(\frac{1}{6}\left(n-n^{-1}\right)+\frac{\alpha^{2}}{2\pi^{2}n}\right)\ln(m\epsilon), & \text{for Dirac fermion},\\
\ln Z_{n}^{(0)}(\alpha) & =\left(\frac{1}{3}\left(n-n^{-1}\right)-\frac{\alpha^{2}}{2\pi^{2}n}+\frac{\left|\alpha\right|}{\pi n}\right)\ln(m\epsilon), & \text{for complex boson}.
\end{align}
The $\ell$-independent contributions of order $O(1)$ (in the limit
$\epsilon\rightarrow0$) are neglected explicitly and correspond to
non-universal quantities.

From Eq. \eqref{eq:LogZnU(1)Approx} the symmetry resolved von Neumann
($n=1$) and Rényi entropies can be straightforwardly computed following
\cite{mdc-20b}. We only report here the leading and sub-leading contributions
to the entropies as 
\begin{equation}
\begin{split}S_{n}(q_{A})-S_{n}\simeq & -\frac{1}{2}\ln\ln\left(m\varepsilon\right)^{-1}+\mathcal{O}(1)\text{ for Dirac fermion}\,,\\
S_{n}(q_{A})-S_{n}\simeq & -\ln\ln\left(m\varepsilon\right)^{-1}+\mathcal{O}(1)\text{ for complex boson}\,,
\end{split}
\label{snqf}
\end{equation}
for $n\geq1$. The total Rényi and von Neumann entropy has the following
large $\ell$ behaviour 
\begin{equation}
\begin{split} & S_{n}\simeq\frac{c}{6}\left(n-n^{-1}\right)\log(m\varepsilon)^{-1}-\frac{n}{n-1}\frac{e^{-2m\ell}}{4\pi m\ell}+\dots\quad n>1,\\
 & S_{1}\simeq\frac{c}{3}\log(m\varepsilon)^{-1}-\frac{1}{4}K_{0}(2m\ell)+\dots,
\end{split}
\end{equation}
where the non-universal constants are again neglected explicitly and
$c=1,2$ for complex fermions and bosons, respectively. We recall
that the result for $S_{1}$ is already known from Ref. \cite{ccd-08};
in particular we get $-\frac{1}{4}K_{0}(2m\ell)$ which is the double
of what was found in the Ising field theory because two particles are present in the
theory. In contrast, for $n>1$ the $n$-th entropy depends strongly
on the theory under consideration beyond the leading order reported
above.

At leading order in the limit $\epsilon\rightarrow0$, one observes
the equipartition of entanglement, namely that $S_{n}(q_{A})$ does
not depend explicitly on $q_{A}$. Nevertheless, the equipartition
is broken explicitly whenever $\varepsilon\neq0$. A careful analysis
has been already performed in \cite{mdc-20b}, where it has been shown
that the terms which break explicitly equipartition maybe be written
as a power series in $\frac{1}{\log(m\varepsilon)^{-1}}$.

Finally we want to mention that the total von Neumann entropy can be written as \cite{ScPlusSf} 
\begin{equation}
S=\sum_{q_{A}}p(q_{A})S(q_{A})-\sum_{q_{A}}p(q_{A})\ln p(q_{A})=S^{c}+S^{f}\,,\label{eq:SfSc}
\end{equation}
where $p(q_{A})=\mathcal{Z}_{1}(q_{A})$ equals the probability
of finding $q_{A}$ as the outcome of a measurement of $\hat{Q}_{A}$.
The contribution $S^{c}$ is called the configurational entanglement
entropy and measures the total entropy due to each charge sector (weighted
with the corresponding probability) \cite{Greiner,Wiseman} and $S^{f}$
denotes the fluctuation (or number) entanglement entropy, which instead
takes into account the entropy due to the fluctuations of the value
of the charge in the subsystem $A$ \cite{Greiner,FreeF3,MBL,kufs-20c}.
In Eq. \eqref{snqf} the log-log term is necessary in the SR quantity to cancel the same contribution to the total entropy coming from $S_f$.

\section{Conclusions\label{sec:Conclusions}}

In this paper we applied the 1+1D bootstrap approach to compute the
form factors of the composite branch-point twist fields which are
directly related to symmetry resolved entropies. The technique was
initiated in Ref. \cite{Z2IsingShg} for discrete symmetries and we
generalised here to a $U(1)$ conserved charge. For simplicity, we
focused on free theories, namely the free massive Dirac theory and
the free massive complex boson theory, both of which admits a $U(1)$
symmetry. The motivation for the study of these free theories is twofold.
In fact, despite the absence of interactions, the form factors of
the corresponding composite branch-point twist fields are highly non-trivial.
As a consequence, these calculations serve both as a warm-up toward
interacting theories (without many technical complications like non-diagonal
scattering etc.) and as a reference point to test future results for integrable models
in the non-interacting limit.

We determined all the form factors of the composite field using two
different methods: standard form factor bootstrap approach and diagonalisation
in replica space. For the Dirac theory, form factors with higher particle
numbers turned out to be Pfaffians of the 2-particle ones. For the
complex boson, form factors with many particles can be obtained using
Wick's theorem and the 2-particle form factor. The form factors for
odd particle numbers and for non-neutral (with respect to the $U(1)$
charge) particle configurations are zero. Our solutions have been
tested using the $\Delta$-theorem sum rule. Furthermore, we recovered
known particular limits: when the flux corresponding to the $U(1)$
charge is zero, the conventional branch-point twist field form factors
are re-obtained, whereas for a single replica we got the conventional
$U(1)$ twist field from factors. In addition, we also determined
the exact vacuum expectation values for the composite fields. 

Although the main goal of this work was the complete determination
of the non-trivial form factors, we also presented the asymptotic
results for the charged moments, the charged partition functions,
and, eventually, the symmetry resolved entropies. We used the two-particle
approximation for the 2-point correlation function of the composite
fields, which gives accurate results for the charged moments when
the considered subsystem is a long interval embedded in the ground state of an infinite system. 
For $n=1$, a non-trivial analytic continuation
is also necessary. We showed that the charged moments obtained from
form factor sums equal those in Ref. \cite{mdc-20b} obtained by
completely different means. Following Ref. \cite{mdc-20b}, we finally
re-derived the charged partition functions and the symmetry resolved
entropies, including sub-leading corrections that breaks the equipartition
of entanglement.

Our results are the starting point to systematically compute higher
and higher corrections to the symmetry resolved entanglement and to
consider subsystems consisting of disjoint intervals as well. These
directions can be an interesting object of future studies. Finally,
the application of these form factor techniques to interacting integrable
field theories with $U(1)$ symmetry is on the way and will be presented
in forthcoming publications.

\subsection*{Acknowledgments}

We are grateful to Giuseppe Di Giulio, Cecilia De Fazio, and Olalla Castro-Alvaredo
for useful discussions. All authors acknowledge support from ERC under
Consolidator grant number 771536 (NEMO).

\appendix

\section{Analytic continuation for $f_{D/B}^{\alpha}(\vartheta,n)$ \label{sec:AppendixA-AnalContForfDB}}

The analytic continuation of the quantity $f_{\kappa}^{\alpha}(\vartheta,n)$
with $\kappa=D,B$ is a subtle issue. The corresponding quantities
$f(\vartheta,n)$ for the standard branch-point twist field of the
Ising and sinh-Gordon models were carefully analysed in Ref. \cite{ccd-08}
where it was shown that the analytic continuation $\tilde{f}(\vartheta,n)$
with domain $n\in[1,\infty)$ can be defined from $f(\vartheta,n)$
for $n=2,3,...$. It turned out that $\tilde{f}(\vartheta,n)=f(\vartheta,n)$
for integer $n\geq2$, but, for $n\rightarrow1$, $f(\vartheta,1)=0$
everywhere except in the origin, where it converges to $\frac{1}{2}$.
(To be more precise, $\tilde{f}(0,n)$ as well as $f(0,n)$ are understood
as $\lim_{\vartheta\rightarrow0}\tilde{f}(\vartheta,n)$ and $\lim_{\vartheta\rightarrow0}f(\vartheta,n)$.
The non-uniform convergence is present in $\lim_{n\rightarrow1}\lim_{\vartheta'\rightarrow\vartheta}\tilde{f}(\vartheta',n)$
as a function of $\vartheta$.) The convergence in
$n$ is therefore non-uniform, which results in a $\delta$-function
in the derivative ${\displaystyle \lim_{n\rightarrow1}\frac{\partial}{\partial n}\tilde{f}(\vartheta,n)}$,
yielding 
\begin{equation}
\lim_{n\rightarrow1}\frac{\partial}{\partial n}\tilde{f}(\vartheta,n)=\pi^{2}\frac{1}{2}\delta(\vartheta)\,.\label{eq:FAnalContIsing}
\end{equation}
This derivative is an important quantity as it governs the leading
long-distance behaviour of the entanglement entropy. However, not
all the analysis of Ref. \cite{ccd-08} is necessary here to obtain
$\tilde{f}_{\kappa}^{\alpha}(\vartheta,n)$: only some of the basic
ideas from \cite{ccd-08} are needed and presented here accordingly.

We first claim, that for the conventional branch-point twist field,
the functions $\tilde{f}_{\kappa}(\vartheta,n)$ can be simply written
as 
\begin{equation}
\tilde{f}_{\kappa}(\vartheta,n)=-\frac{\tanh\frac{\vartheta}{2}}{\langle\mathcal{T}_{D/B,n}\rangle^{2}}2\text{Im}\left[F_{2}^{\mathcal{T}_{\kappa}|(+1)(-1)}(-2\vartheta+i\pi,n)-F_{2}^{\mathcal{T}_{\kappa}|(+1)(-1)}(-2\vartheta+2in\pi-i\pi,n)\right]\,.\label{fTildeDB}
\end{equation}
This can be justified following the logic of \cite{ccd-08}: recalling
the definition of $\tilde{f}_{D/B}(\vartheta,n)$ 
\begin{equation}
\langle\mathcal{T}_{\kappa,n}\rangle^{2}f_{\kappa}(\vartheta,n)=2\sum_{j=0}^{n-1}F_{2}^{\mathcal{T}_{\kappa}|(+1)(-1)}(-\vartheta+2\pi ij)\left(F_{2}^{\mathcal{T}_{\kappa}|(+1)(-1)}(-\vartheta+2\pi ij)\right)^{*}=\sum_{j=0}^{n-1}s_{\kappa}(\vartheta,j),
\end{equation}
one can consider the following contour integral and treat $j$ as
a continuous parameter 
\begin{equation}
\frac{1}{2\pi i}\oint_{\mathcal{C}}\text{d}z\pi\cot(\pi z)s_{\kappa}(\vartheta,z)\,,\label{cint2-1}
\end{equation}
where the contour is a rectangle with vertices $(-\epsilon-iL,n-\epsilon-iL,n-\epsilon+iL,-\epsilon+iL)$.
This contour integral is zero as when $L\rightarrow\infty$, the contributions
of the horizontal lines vanish and, at least in free field theories,
the vertical contributions cancel each other due to the periodicity
of $s_{\kappa}(\vartheta,z+n)=S_{\kappa}^{2}s_{\kappa}(\vartheta,z)$
and $S_{\kappa}^{2}=1$ for $\kappa=D,B$. We can evaluate the integral
\eqref{cint2-1} by the residue theorem; the poles are at the positions
$z=1,2,\ldots,n-1$, at $z=\frac{1}{2}\pm\frac{\vartheta}{2\pi i}$,
and $z=n-\frac{1}{2}\pm\frac{\vartheta}{2\pi i}$. Using the explicit
values of the residues, we end up with 
\begin{equation}
0=\sum_{j=0}^{n-1}s_{\kappa}(\vartheta,j)+\frac{\tanh\frac{\vartheta}{2}}{\langle\mathcal{T}_{\kappa,n}\rangle^{2}}2\text{Im}\left[F_{2}^{\mathcal{T}_{\kappa}|(+1)(-1)}(-2\vartheta+i\pi,n)-F_{2}^{\mathcal{T}_{\kappa}|(+1)(-1)}(-2\vartheta+2in\pi-i\pi,n)\right]\,,\label{fTildeDB-1}
\end{equation}
from which Eq. \eqref{fTildeDB} is inferred.

More importantly, the derivatives of $\tilde{f}_{D}(\vartheta,n)$
and $\tilde{f}_{B}(\vartheta,n)$ at $n=1$ have essentially the same
$\delta$-type behaviour. It is straightforward to show that the difference
of the two functions can be recast as 
\begin{equation}
\tilde{f}_{D}(\vartheta,n)-\tilde{f}_{B}(\vartheta,n)=\frac{4\tanh\frac{\vartheta}{2}\cos^{2}\frac{\pi}{2n}\cosh\frac{\vartheta}{n}\tanh\frac{\vartheta}{2n}\left(2\cosh\frac{\vartheta}{n}+\cosh\frac{2\vartheta}{n}-2\cos\frac{\pi}{n}+3\right)}{n\left(\cosh\frac{2\vartheta}{n}-\cos\frac{2\pi}{n}\right)}\,.\label{fD-fB}
\end{equation}
This is a smooth analytic function also when $n$ approaches one.
Differentiating \eqref{fD-fB} and taking $n\to1$, we obtain identically
zero. Therefore, combining Eq. \eqref{eq:FAnalContIsing} with $\tilde{f}_{D}(\vartheta,n)=2\tilde{f}_{\text{Ising}}(\vartheta,n)$,
we have 
\begin{equation}
\begin{split}\lim_{n\rightarrow1}\tilde{f}_{\kappa}(\vartheta,n)= & \begin{cases}
0 & \vartheta\neq0,\\
1 & \vartheta=0.
\end{cases}\end{split}
\label{AnConN1DB}
\end{equation}
\begin{equation}
\begin{split}\lim_{n\rightarrow1}\frac{\partial}{\partial n}\tilde{f}_{\kappa}(\vartheta,n)= & \pi^{2}\delta(\vartheta)\,.\end{split}
\end{equation}

In a similar spirit, we can easily derive the corresponding derivatives
for $\tilde{f}_{\kappa}^{\alpha}(\vartheta,n)$. For both the complex
boson and the Dirac theory, we can write $\tilde{f}_{\kappa}^{\alpha}(\vartheta,n)$
as 
\begin{multline}
\langle\mathcal{T}_{\kappa,n}^{\alpha}\rangle^{2}f_{\kappa}^{\alpha}(\vartheta,n)=\sum_{j=0}^{n-1}F_{2}^{\mathcal{T}_{\kappa}^{\alpha}|(+1)(-1)}(-\vartheta+2\pi ij)\left(F_{2}^{\mathcal{T}_{\kappa}^{\alpha}|(+1)(-1)}(-\vartheta+2\pi ij)\right)^{*}+\\
\sum_{j=0}^{n-1}F_{2}^{\mathcal{T}_{\kappa}^{\alpha}|(-1)(+1)}(-\vartheta+2\pi ij)\left(F_{2}^{\mathcal{T}_{\kappa}^{\alpha}|(-1)(+1)}(-\vartheta+2\pi ij)\right)^{*}=\sum_{j=0}^{n-1}s_{\kappa}^{\alpha}(\vartheta,j),
\end{multline}
and consider the contour integral 
\begin{equation}
\frac{1}{2\pi i}\oint_{\mathcal{C}}\text{d}z\pi\cot(\pi z)s_{\kappa}^{\alpha}(\vartheta,z)\,,\label{cint2}
\end{equation}
with the same contour as before. This contour integral is again zero
for the same reasons as above: the contributions of the horizontal lines
vanish and, in free field theories, the vertical contributions cancel
each other because $S_{\kappa}^{2}=1$. We can evaluate the integral
\eqref{cint2} by the residue theorem. The poles are at the positions
$z=1,2,\ldots,n-1$, at $z=\frac{1}{2}\pm\frac{\vartheta}{2\pi i}$,
and $z=n-\frac{1}{2}\pm\frac{\vartheta}{2\pi i}$. Using the explicit
values of the residues, we end up with 
\begin{equation}
\begin{split}\sum_{j=1}^{n-1}s_{\kappa}^{\alpha}(\vartheta,j)= & -\frac{\tanh\frac{\vartheta}{2}}{\langle\mathcal{T}_{\kappa,n}^{\alpha}\rangle^{2}}\text{Im}\left[F_{2}^{\mathcal{T}_{\kappa}^{\alpha}|(+1)(-1)}(-2\vartheta+i\pi,n)+F_{2}^{\mathcal{T}_{\kappa}^{\alpha}|(-1)(+1)}(-2\vartheta+i\pi,n)\right.\\
 & \left.-e^{-i\alpha}F_{2}^{\mathcal{T}_{\kappa}^{\alpha}|(+1)(-1)}(-2\vartheta+i2\pi n-i\pi,n)-e^{i\alpha}F_{2}^{\mathcal{T}_{\kappa}^{\alpha}|(-1)(+1)}(-2\vartheta+i2\pi n-i\pi,n)\right]
\end{split}
\end{equation}
from which the analytic continuation of $f_{\kappa}^{\alpha}(\vartheta,n)$
for both $\kappa=D,B$ is inferred as 
\begin{equation}
\begin{split}\tilde{f}_{\kappa}^{\alpha}(\vartheta,n)= & -\frac{\tanh\frac{\vartheta}{2}}{\langle\mathcal{T}_{\kappa,n}^{\alpha}\rangle^{2}}\text{Im}\left[F_{2}^{\mathcal{T}_{\kappa}^{\alpha}|(+1)(-1)}(-2\vartheta+i\pi,n)+F_{2}^{\mathcal{T}_{\kappa}^{\alpha}|(-1)(+1)}(-2\vartheta+i\pi,n)\right.\\
 & \left.-e^{-i\alpha}F_{2}^{\mathcal{T}_{\kappa}^{\alpha}|(+1)(-1)}(-2\vartheta+i2\pi n-i\pi,n)-e^{i\alpha}F_{2}^{\mathcal{T}_{\kappa}^{\alpha}|(-1)(+1)}(-2\vartheta+i2\pi n-i\pi,n)\right]\,.
\end{split}
\label{fTildeU1BD}
\end{equation}
It is easy to check that $\tilde{f}_{\kappa}^{\alpha}(\vartheta,n)=f_{\kappa}^{\alpha}(\vartheta,n)$
for integers $n>1$.

The derivative of $\tilde{f}_{\kappa}^{\alpha}(\vartheta,n)$ can
be obtained without further work exploiting the property that the
function $\tilde{f}_{D/B}^{\alpha}(\vartheta,n)-\cos\alpha\,\tilde{f}_{D/B}(\vartheta,n)$
is smooth and converges to a smooth function as $n\rightarrow1$.
Indeed, using Eqs. \eqref{fTildeDB} and \eqref{fTildeU1BD} we immediately
have 
\begin{equation}
\tilde{f}_{D}^{\alpha}(\vartheta,1)-\cos\alpha\,\tilde{f}_{D}(\vartheta,1)=\frac{8\sinh^{2}\vartheta\sin^{2}\frac{\alpha}{2}\cosh\frac{\vartheta\alpha}{\pi}}{(\cosh\vartheta+1)(\cosh(2\vartheta)-1)}\,,
\end{equation}
and 
\begin{equation}
\begin{split}\lim_{n\rightarrow1}\frac{\partial}{\partial n}\left[\tilde{f}_{D}^{\alpha}(\vartheta,n)-\cos\alpha\,\tilde{f}_{D}(\vartheta,n)\right]= & \frac{1}{\pi(\cosh\vartheta+1)}\left[2\pi\cosh\frac{\vartheta\alpha}{\pi}\left(2\vartheta\coth\vartheta\sin^{2}\frac{\alpha}{2}-2\sin^{2}\frac{\alpha}{2}-\alpha\sin\alpha\right)\right.\\
 & \left.\qquad+2\sinh\frac{\vartheta\alpha}{\pi}\left(\pi^{2}\coth\vartheta\sin\alpha-2\vartheta\alpha\sin^{2}\frac{\alpha}{2}\right)\right]
\end{split}
\end{equation}
for the Dirac theory and 
\begin{equation}
\tilde{f}_{B}^{\alpha}(\vartheta,1)-\cos\alpha\,\tilde{f}_{B}(\vartheta,1)=2\text{sech}^{2}\frac{\vartheta}{2}\sin^{2}\frac{\alpha}{2}\cosh\left(\vartheta-\frac{\vartheta\left|\alpha\right|}{\pi}\right)\,,
\end{equation}
and 
\begin{multline}
\lim_{n\rightarrow1}\frac{\partial}{\partial n}\left[\tilde{f}_{B}^{\alpha}(\vartheta,n)-\cos\alpha\,\tilde{f}_{B}(\vartheta,n)\right]=\frac{2\tanh\frac{\vartheta}{2}\text{csch}^{2}\vartheta\left|\sin\frac{\alpha}{2}\right|}{\pi}\times\\
\times\left[\pi\cos\frac{\alpha}{2}\left((2\pi-\left|\alpha\right|)|\sinh\frac{\vartheta\alpha}{\pi}|-\left|\alpha\right|\sinh\left(\vartheta\left(2-\frac{\left|\alpha\right|}{\pi}\right)\right)\right)\right.\\
\left.\qquad\qquad+\left|\sin\frac{\alpha}{2}\right|\left(\vartheta\left|\alpha\right|\cosh\left(\vartheta\left(2-\frac{\left|\alpha\right|}{\pi}\right)\right)+\vartheta(2\pi-\left|\alpha\right|)\cosh\frac{\vartheta\alpha}{\pi}\right)\right.\\
\left.\qquad\qquad-2\pi\left|\sin\frac{\alpha}{2}\right|\sinh\vartheta\cosh\left(\vartheta-\frac{\vartheta\left|\alpha\right|}{\pi}\right)\right]
\end{multline}
for the complex boson theory.

These expressions lead to the main result of this appendix namely
\begin{equation}
\begin{split}\lim_{n\rightarrow1}\tilde{f}_{D}^{\alpha}(\vartheta,n)= & \begin{cases}
\frac{8\sinh^{2}\vartheta\sin^{2}\frac{\alpha}{2}\cosh\frac{\vartheta\alpha}{\pi}}{(\cosh\vartheta+1)(\cosh(2\vartheta)-1)} & \vartheta\neq0,\\
1 & \vartheta=0,
\end{cases}\end{split}
\label{AnConN1Dirac}
\end{equation}
and 
\begin{equation}
\begin{split}\lim_{n\rightarrow1}\frac{\partial}{\partial n}\tilde{f}_{D}^{\alpha}(\vartheta,n)= & \pi^{2}\cos\alpha\,\delta(\vartheta)+\frac{1}{\pi(\cosh\vartheta+1)}\left[2\pi\cosh\frac{\vartheta\alpha}{\pi}\left(2\vartheta\coth\vartheta\sin^{2}\frac{\alpha}{2}-2\sin^{2}\frac{\alpha}{2}-\alpha\sin\alpha\right)\right.\\
 & \left.\qquad\qquad\qquad\qquad\qquad+2\sinh\frac{\vartheta\alpha}{\pi}\left(\pi^{2}\coth\vartheta\sin\alpha-2\vartheta\alpha\sin^{2}\frac{\alpha}{2}\right)\right],
\end{split}
\label{AnConDerivativeDirac}
\end{equation}
for the Dirac theory and 
\begin{equation}
\begin{split}\lim_{n\rightarrow1}\tilde{f}_{B}^{\alpha}(\vartheta,n)= & \begin{cases}
2\text{sech}^{2}\frac{\vartheta}{2}\sin^{2}\frac{\alpha}{2}\cosh\left(\vartheta-\frac{\vartheta\left|\alpha\right|}{\pi}\right) & \vartheta\neq0,\\
1 & \vartheta=0,
\end{cases}\end{split}
\label{AnConN1Boson}
\end{equation}
and 
\begin{multline}
\lim_{n\rightarrow1}\frac{\partial}{\partial n}\tilde{f}_{B}^{\alpha}(\vartheta,n)=\pi^{2}\cos\alpha\,\delta(\vartheta)+\frac{2\tanh\frac{\vartheta}{2}\text{csch}^{2}\vartheta|\sin\frac{\alpha}{2}|}{\pi}\times\\
\times\left[\pi\cos\frac{\alpha}{2}\left((2\pi-\left|\alpha\right|)\left|\sin\frac{\alpha}{2}\right|-\left|\alpha\right|\sinh\left(\vartheta\left(2-\frac{\left|\alpha\right|}{\pi}\right)\right)\right)\right.\\
\left.\qquad\qquad\qquad\qquad\qquad+\left|\sin\frac{\alpha}{2}\right|\left(\vartheta\left|\alpha\right|\cosh\left(\vartheta\left(2-\frac{\left|\alpha\right|}{\pi}\right)\right)+\vartheta(2\pi-\left|\alpha\right|)\cosh\frac{\vartheta\alpha}{\pi}\right)\right.\\
\left.\qquad\qquad\qquad\qquad\qquad-2\pi\left|\sin\frac{\alpha}{2}\right|\sinh\vartheta\cosh\left(\vartheta-\frac{\vartheta\left|\alpha\right|}{\pi}\right)\right],\label{AnconDerivativeBoson}
\end{multline}
for the complex boson.

The computation of further, sub-leading $\ell$-dependent corrections
is not addressed in this work, but can be straightforwardly achieved.
To use Eq. \eqref{eq:BesselTheta2n}, one eventually needs to expand
$\tilde{f}_{\kappa}^{\alpha}(\vartheta,n)$, $\tilde{f}_{\kappa,analytic}^{\alpha}(\vartheta,1)$
and $\lim_{n\rightarrow1}\frac{\partial}{\partial n}\tilde{f}_{\kappa,analytic}^{\alpha}(\vartheta,n)$
into series around $\vartheta=0$, which can be computed easily using
the above formulas. The quantities $\tilde{f}_{\kappa,analytic}^{\alpha}(\vartheta,1)$
and $\lim_{n\rightarrow1}\frac{\partial}{\partial n}\tilde{f}_{\kappa,analytic}^{\alpha}(\vartheta,n)$
are understood as first performing the $n\rightarrow1$ limit and
only then the $\vartheta'\rightarrow\vartheta$ limit (and in particular
$\vartheta'\rightarrow0$). For the Dirac fermions we have 
\begin{equation}
\tilde{f}_{D}^{\alpha}(\vartheta,n)=1+\frac{-\pi^{2}\left(n^{2}+2\right)+12\pi\csc\frac{\pi}{n}\left(\alpha\sin\frac{\alpha}{n}+\pi\cot\frac{\pi}{n}\cos\frac{\alpha}{n}\right)+6\alpha^{2}}{12\pi^{2}n^{2}}\vartheta^{2}+\mathcal{O}(\vartheta^{4}),\label{AnConFTildenDirac}
\end{equation}
for $n\neq1$ and 
\begin{equation}
\tilde{f}_{D,analytic}^{\alpha}(\vartheta,1)=2\sin^{2}\frac{\alpha}{2}-\frac{\left(\left(\pi^{2}-2\alpha^{2}\right)\sin^{2}\frac{\alpha}{2}\right)}{2\pi^{2}}\vartheta^{2}+\mathcal{O}(\vartheta^{4}),\label{fTildeDiracn1AnalyticPart}
\end{equation}
for $n=1$ and for the derivative 
\begin{equation}
\lim_{n\rightarrow1}\frac{\partial}{\partial n}\tilde{f}_{D,analytic}^{\alpha}(\vartheta,n)=\left(-\frac{\alpha^{3}\sin\alpha}{3\pi^{2}}-\frac{2\alpha^{2}\sin^{2}\frac{\alpha}{2}}{\pi^{2}}+\frac{2}{3}\sin^{2}\frac{\alpha}{2}+\frac{1}{3}\alpha\sin\alpha\right)\vartheta^{2}+\mathcal{O}(\vartheta^{4})\,,\label{DerivativefTildeDiracAnalyticpart}
\end{equation}
and for the complex Boson 
\begin{equation}
\tilde{f}_{B}^{\alpha}(\vartheta,n)=1-\frac{\pi^{2}\left(n^{2}-4\right)+6\pi\csc^{2}\frac{\pi}{n}\left(\left|\alpha\right|\cos\frac{2\pi-\left|\alpha\right|}{n}+(2\pi-\left|\alpha\right|)\cos\frac{\alpha}{n}\right)-6\alpha^{2}+12\pi\left|\alpha\right|}{12\pi^{2}n^{2}}\vartheta^{2}+\mathcal{O}(\vartheta^{4})\label{AnConFTildenBoson}
\end{equation}
for $n\neq1$ and 
\begin{equation}
\tilde{f}_{B,analytic}^{\alpha}(\vartheta,1)=2\sin^{2}\frac{\alpha}{2}+2\left(\frac{1}{2}\left(1-\frac{\left|\alpha\right|}{\pi}\right)^{2}-\frac{1}{4}\right)\sin^{2}\frac{\alpha}{2}\,\vartheta^{2}+\mathcal{O}(\vartheta^{4}),\label{fTildeBosonAnalyticpart}
\end{equation}
for $n=1$ and for the derivative 
\begin{multline}
\lim_{n\rightarrow1}\frac{\partial}{\partial n}\tilde{f}_{B,analytic}^{\alpha}(\vartheta,n)=\\
=-\left(\frac{\left|\alpha\right|(\left|\alpha\right|-2\pi)(\left|\alpha\right|-\pi)|\sin\alpha|-\left(3\alpha^{2}-6\pi\left|\alpha\right|+2\pi^{2}\right)(\cos\alpha-1)}{3\pi^{2}}\right)\vartheta^{2}+\mathcal{O}(\vartheta^{4})\,.\label{DerivativefTildeBosonAnalyticpart}
\end{multline}

We conclude this appendix mentioning the behaviour for $n\rightarrow\infty$,
for which we are going to show that the limiting functions for $\tilde{f}_{\kappa}^{\alpha}(\vartheta,n)$
and $\tilde{f}_{\kappa}(\vartheta,n)$ are the same. More precisely,
we have that 
\begin{equation}
\lim_{n\rightarrow\infty}\tilde{f}_{D}^{\alpha}(\vartheta,e^{i\phi}n+c)=2\frac{\left(2\vartheta^{2}+\pi^{2}\right)\tanh\frac{\vartheta}{2}}{\vartheta\left(\vartheta^{2}+\pi^{2}\right)}\,,
\end{equation}
and 
\begin{equation}
\lim_{n\rightarrow\infty}\tilde{f}_{B}^{\alpha}(\vartheta,e^{i\phi}n+c)=2\frac{\pi^{2}\tanh\frac{\vartheta}{2}}{\vartheta\left(\vartheta^{2}+\pi^{2}\right)}\,,
\end{equation}
for any constant $c$ and any direction $\phi$ on the complex plane.
This large $n$ behaviour is related to the unicity of the analytic
continuation \cite{ccd-08} by Carlson's theorem \cite{Carlson}.
Indeed, let us suppose the existence of another function $\tilde{g}_{\kappa}^{\alpha}(\vartheta,n)$,
which satisfies $\tilde{g}_{\kappa}^{\alpha}(\vartheta,n)=\tilde{f}_{\kappa}^{\alpha}(\vartheta,n)$
for $n$-s with $n\geq1$. We assume that $\tilde{g}_{\kappa}^{\alpha}(\vartheta,n)|<Ce^{q|n|}$
for $\text{Re}(n)>0$ and with $q<\frac{\pi}{2}$; this assumption
is motivated by the fact that both $\text{Tr}\left(\rho_{A}^{n}\right)$
and $\text{Tr}\left(\rho_{A}^{n}e^{i\alpha\hat{Q}_{A}}\right)$ behave
so for finite systems, see again Ref. \cite{ccd-08} for a detailed
discussion. Then Carlson's theorem can be applied to $\tilde{f}_{\kappa}^{\alpha}(\vartheta,n)-\tilde{g}_{\kappa}^{\alpha}(\vartheta,n)$
, which is zero for all positive integers. Based on our motivating
assumptions for $\tilde{g}_{\kappa}^{\alpha}(\vartheta,n)$ and the
actual behaviour of $\tilde{f}_{\kappa}^{\alpha}(\vartheta,n)$ the
theorem implies the unicity of the continuation $\tilde{f}_{\kappa}^{\alpha}(\vartheta,n)$.

\section{Vacuum expectation value of $\mathcal{T}_{D.n}^{\alpha}$\label{sec:AppendixB-VEVDirac}}

The determination of the vacuum expectation value (VEV), i.e., the
zero particle FF is generally a difficult task and for the branch-point
twist fields, the VEV has been derived only for the Ising model, both
for the conventional field \cite{ccd-08} and for the composite one
\cite{Z2IsingShg} and for the conventional twist
field in the complex boson theory \cite{DoyonBosonU1VEV}. 
Strictly speaking  the VEV is not unique, of course, but prescribing the standard
conformal normalisation, i.e., requiring that
\begin{equation}
\lim_{l\rightarrow0}\langle\mathcal{T}_{n}^{\alpha}(\ell,0)\tilde{\mathcal{T}}_{n}^{\alpha}(0,0)\rangle\left|\ell\right|^{2d_{n}^{\alpha}}=1
\end{equation}
it can be usually unambiguously defined. In this
appendix, we derive the VEV $\mathcal{T}_{D,n}^{\alpha}$ for the
Dirac field using and modifying ideas taken from Refs. \cite{ccd-08,cs-17,CFH,Z2IsingShg}.
Following these ideas, we eventually proceed in a
very similar way to the logic of section \ref{secAlternative}, that
is, we essentially perform a diagonalisation in the space of replicas
and exploit the factorisation of the field into Fourier components in the replica space. 

Let us consider Dirac fermion fields and denote the one living on
the $j$th replica by $\Psi_{j}$. Similarly to section \ref{secAlternative},
we can consider the replica space spanned by vectors of the form $\left(\Psi_{1}^{\text{}},...,\Psi_{n}^{\text{}}\right)^{T}$
and search for a matrix $\tau$ whose action in the space replica
corresponds to the $U(1)$ composite twist field. Since the total
phase picked up by the twist field when turning around the entire
Riemann surface is $e^{i\alpha}$, the transformation matrix $\tau$
can be obtained by multiplying the transformation matrix of the conventional fields  \cite{CFH} by $e^{i\alpha/n}$, as done in Ref. \cite{mdc-20b}. Accordingly
the following representation is obtained
\begin{equation}
\tau=e^{i\frac{\alpha}{n}}\left(\begin{array}{ccccccc}
0 & 0 & 0 & 0 & \cdots & 0 & (-1)^{n+1}\\
1 & 0 & 0 & 0 & \cdots & 0 & 0\\
0 & 1 & 0 & 0 & \cdots & 0 & 0\\
0 & 0 & 1 & 0 &  & 0 & 0\\
\vdots & \vdots &  & \ddots & \ddots &  & \vdots\\
0 & 0 & 0 & 0 & \ddots & 0 & 0\\
0 & 0 & 0 & 0 & \cdots & 1 & 0
\end{array}\right) \,.\label{eq:sigmaMatrixCassiniHuerta(-1)}
\end{equation}
The transformation matrix $\tau$ has to be diagonalised for
the determination of the VEV \cite{ccd-08}. The eigenvalues of $\tau$
can be written as $e^{i2\pi k/n}e^{i\alpha/n}$ with $k$ ranging
from $-(n-1)/2$ to $(n-1)/2$ for both even and odd $n$. Using the
eigenvectors of $\tau$ new fermion fields $\Psi_{k}$ can be defined
as well. These new fermion fields satisfy the canonical anticommutation
relations $\{\Psi_{k}(x),\Psi_{k'}^{\dagger}(x')\}=\delta_{k,k'}\delta(x-x')$,
$\{\Psi_{k}(x),\Psi_{k'}(x')\}=0$ and $\{\Psi_{k}^{\dagger}(x),\Psi_{k'}^{\dagger}(x')\}=0$.

The eigenvalues of the transformation $\tau$ obtained above are compatible
with the four-point function of the composite $U(1)$ branch-point
twist field \cite{mdc-20b} 
\begin{equation}
\frac{\langle\Psi_{k}^{\dagger}(z)\Psi_{k}(z')\mathcal{T}_{n}^{\alpha}(w)\tilde{\mathcal{T}}_{n}^{\alpha}(w')\rangle}{\langle\mathcal{T}_{n}^{\alpha}(w)\tilde{\mathcal{T}}_{n}^{\alpha}(w')\rangle}=\frac{1}{z-z'}\left(\frac{\left(z-w\right)\left(z'-w'\right)}{\left(z-w'\right)\left(z'-w\right)}\right)^{\frac{k}{n}+\frac{\alpha}{2\pi n}}\,,\label{eq:ComplexFermionDublet4ptFunctionU1Dirac}
\end{equation}
at the UV critical point. Eq. \eqref{eq:ComplexFermionDublet4ptFunctionU1Dirac}
means that performing a clock-wise turn on $\Psi_{k}(z')$ around
the twist field $\mathcal{T}^{\alpha}$ inserted at $w$, the correct
factor of $e^{i\pi(2k+\alpha)/n}$ is accumulated. From Eq. (\ref{eq:ComplexFermionDublet4ptFunctionU1Dirac})
the factorisation of the composite $U(1)$ branch-point twist field
naturally follows and this fact makes it possible to compute the UV
dimensions of the factorised components and eventually to determine
the VEV in the massive theory. The factorisation of the composite
$U(1)$ branch-point twist field in our case can be written as 
\begin{equation}
\mathcal{T}_{n}^{\alpha}(w)=\prod_{k=-(n-1)/2}^{(n-1)/2}\mathcal{T}_{k,n}^{\alpha}(w)\,,
\end{equation}
and the action of $\mathcal{T}_{k,n}^{\alpha}(w)$ is non trivial
only on the $\Psi_{k}$ and $\Psi_{k}^{\dagger}$ fields. The (chiral)
conformal dimension of $\mathcal{T}_{k,n}^{\alpha}$ can be can be
obtained from the relation \cite{cc-04,k-87,dixon} 
\begin{equation}
\frac{\langle T_{k}(z)\mathcal{T}_{k,n}^{\alpha}(w)\tilde{\mathcal{T}}_{k,n}^{\alpha}(w')\rangle}{\langle\mathcal{T}_{k,n}^{\alpha}(w)\tilde{\mathcal{T}}_{k,n}^{\alpha}(w')\rangle}=\Delta_{k}\frac{\left(w-w'\right)^{2}}{\left(z-w\right)^{2}\left(z-w'\right)^{2}}\,,\label{eq:TTDTDTilde}
\end{equation}
where $T_{k}$ is the stress-energy tensor of the $k$ components,
as the stress-energy tensor can be shown to factorise into $k$ sectors
as well. The analysis of computing the conformal dimensions was performed
in \cite{mdc-20b} leading us to the result 
\begin{equation}
\Delta_{k}=\frac{1}{2}\left(\frac{k}{n}+\frac{\alpha}{2\pi n}\right)^{2}.
\end{equation}

It can now be checked that the total conformal dimension of the composite
twist field agrees with the known value $\frac{1}{24}\left(n-n^{-1}\right)+\frac{\alpha^{2}}{2(2\pi)^{2}n}$
because for even $n$ we have 
\begin{equation}
\sum_{l=1}^{\frac{n}{2}\left(\frac{l-1/2}{n}\right)^{2}}+\frac{\alpha^{2}}{2\left(2\pi\right)^{2}n}=\frac{1}{24}\left(n-n^{-1}\right)+\frac{\alpha^{2}}{2\left(2\pi\right)^{2}n}\,.
\end{equation}
and for odd $n$ 
\begin{equation}
\sum_{l=1}^{\frac{n-1}{2}} \left(\frac{l}{n}\right)^{2}+\frac{\alpha^{2}}{2\left(2\pi\right)^{2}n}=\frac{1}{24}\left(n-n^{-1}\right)+\frac{\alpha^{2}}{2\left(2\pi\right)^{2}n}\,.
\end{equation}

Based on the above analysis, a natural assumption is that the decomposition
of the composite branch-point twist fields can be equivalently written
in terms of conventional $U(1)$ twist fields/vertex operators $\mathcal{V}_{D}^{\alpha_{p,n}}$
as 
\begin{equation}
\mathcal{T}_{n}^{\alpha}(w,\bar{w})=\prod_{k}\mathcal{V}^{\alpha_{k,n}}(w,\bar{w})=\prod_{k=-\frac{n-1}{2}}^{\frac{n-1}{2}}\mathcal{V}_{D}^{\frac{k}{n}+\frac{\alpha}{2\pi n}}(w,\bar{w})
\end{equation}
since the left and right conformal dimension of these fields is exactly
$\frac{1}{2}\left(\frac{k}{n}+\frac{\alpha}{2\pi n}\right)^{2}$.
Under the plausible assumption that this factorisation of the composite
branch-point twist field also holds in the off-critical theory (which
is also justified by section \ref{secAlternative}) its vacuum expectation
value can be obtained by exploiting the results in Ref. \cite{VEV1}
\begin{equation}
\langle\mathcal{V}_{D}^{\varphi}\rangle=\left(\frac{m}{2}\right)^{\varphi^{2}}\frac{1}{G(1-\varphi)G(1+\varphi)}\,,
\end{equation}
where $G(x)$ is the Barnes G-function. Hence, for the $n$-copy Dirac
theory we have 

\begin{equation}
\langle\mathcal{T}_{D,n}^{\alpha}\rangle=\left(\frac{m}{2}\right)^{\left(\frac{n-n^{-1}}{12}+\frac{\alpha^{2}}{(2\pi)^{2}n}\right)}\prod_{k=-\frac{n-1}{2}}^{\frac{n-1}{2}}\frac{1}{G(1-\frac{2k+\alpha/\pi}{2n})G(1+\frac{2k+\alpha/\pi}{2n})}\,.
\end{equation}
Using the integral representation of the Barnes G-function, we can
rewrite the VEV as 
\begin{multline}
\langle\mathcal{T}_{D,n}^{\alpha}\rangle=\left(\frac{m}{2}\right)^{\left(\frac{n-n^{-1}}{12}+\frac{\alpha^{2}}{(2\pi)^{2}n}\right)}\times\\
\times\exp\left[\int_{0}^{\infty}\frac{\mathrm{d}t}{t}\left(\frac{\sinh t\,\cosh\left(\frac{t\alpha}{\pi n}\right)-n\,\text{\ensuremath{\sinh}}\frac{t}{n}}{2\,\text{\ensuremath{\sinh}}\frac{t}{n}\sinh^{2}t}-\left(\frac{n-n^{-1}}{12}+\frac{\alpha^{2}}{(2\pi)^{2}n}\right)e^{-2t}\right)\right]\,.\label{vev1}
\end{multline}

\section{Vacuum expectation value of $\mathcal{T}_{B,n}^{\alpha}$\label{sec:AppendixC-VEVBoson}}

To determine the VEV of the composite $U(1)$ branch-point twist field
in the free complex boson theory, we can proceed in a similar fashion,
and our eventual computation boils down again to writing the VEV of
the composite branch-point twist field as a product of VEVs for conventional
$U(1)$ twist fields. Contrary to the case of the free Dirac theory,
however, we now face an important subtlety when defining the VEVs
$\langle\mathcal{V}_{B}^{\varphi}\rangle$ in the complex boson theory.
This theory is not compact and as a consequence the short-distance behaviour of the theory and the proper definition of the VEV are non-trivial
 \cite{DoyonBosonU1VEV}, because of the presence of a zero mode. 
Consequently, the explicit value of the VEV is non-universal, in the sense that depends on the employed normalisation.  
This problem was carefully discussed in Ref. \cite{DoyonBosonU1VEV}, where
an expression for the VEV was actually proposed based on a natural
regularisation of the fields and on the angular quantisation scheme.
In the following derivation, we adopt this convention, which was already
used in \cite{DoyonBosonU1VEV} to derive the VEV of the conventional
branch-point twist field (in a similar calculation to what follows).

To proceed with our eventual derivation in the standard
way, we first of all need again the transformation matrix $\tau$,
whose action in the space replica space (i.e. on the vector $\left(\Phi_{1}^{\text{}},...,\Phi_{n}^{\text{}}\right)^{T}$)
corresponds to the composite twist field can now be written as \cite{mdc-20b}
\begin{equation}
\tau=e^{i\frac{\alpha}{n}}\left(\begin{array}{ccccccc}
0 & 0 & 0 & 0 & \cdots & 0 & 1\\
1 & 0 & 0 & 0 & \cdots & 0 & 0\\
0 & 1 & 0 & 0 & \cdots & 0 & 0\\
0 & 0 & 1 & 0 &  & 0 & 0\\
\vdots & \vdots &  & \ddots & \ddots &  & \vdots\\
0 & 0 & 0 & 0 & \ddots & 0 & 0\\
0 & 0 & 0 & 0 & \cdots & 1 & 0
\end{array}\right)\,.\label{eq:sigmaBoson}
\end{equation}

The eigenvalues of $\tau$ are the roots of unity times $e^{i\alpha/n}$
that is $e^{i2\pi k/n}e^{i\alpha/n}$ with $k$ ranging from $0$
to $n-1$. Similarly to the Dirac case, one can introduce new Bose
fields, which satisfy the canonical commutation relations $[\Phi_{k}(x),\Phi_{k'}^{\dagger}(x)]=\delta_{k,k'}\delta(x-x')$,
$[\Phi_{k}(x),\Phi_{k'}(x')]=0$ and $[\Phi_{k}^{\dagger}(x),\Phi_{k'}^{\dagger}(x')]=0$.
In complete analogy with the Dirac case, the factorisation of composite
twist field is inferred as \cite{mdc-20b} 
\begin{equation}
\mathcal{T}_{n}^{\alpha}(w)=\prod_{k=0}^{n-1}\mathcal{T}_{k,n}^{\alpha}(w)\,,
\end{equation}
where the conformal dimension of each component is 
\begin{equation}
\Delta_{k}=\frac{1}{2}\left(\frac{k}{n}+\frac{\left|\alpha\right|}{2\pi n}\right)\left(1-\frac{k}{n}-\frac{\left|\alpha\right|}{2\pi n}\right).
\end{equation}
The total conformal dimension of the composite twist field agrees
with the known value for any $n$, that is 
\begin{equation}
\Delta=\sum_{k=0}^{n}\frac{1}{2}\left(\frac{k}{n}+\frac{\left|\alpha\right|}{2\pi n}\right)\left(1-\frac{k}{n}-\frac{\left|\alpha\right|}{2\pi n}\right)=\frac{1}{12}\left(n-n^{-1}\right)-\frac{\alpha^{2}}{2(2\pi)^{2}n}+\frac{\left|\alpha\right|}{4\pi n}\,.
\end{equation}

Based on the above analysis, the decomposition of the composite branch-point
twist fields is assumed again and is rephrased as 
\begin{equation}
\mathcal{T}_{n}^{\alpha}(w,\bar{w})=\prod_{k=0}^{n}\mathcal{T}_{k,n}^{\alpha}(w,\bar{w})=\prod_{k=0}^{n}\mathcal{V}_{B}^{\varphi_{k}}(w,\bar{w}),
\end{equation}
in terms of conventional bosonic $U(1)$ twist field $\mathcal{V}_{B}^{\varphi_{k}}$, with $\varphi_{k}=\frac{k}{n}+\frac{\alpha}{2\pi n}$
and whose left and right conformal dimensions are exactly $\frac{1}{2}\left(\frac{k}{n}+\frac{\alpha}{2\pi n}\right)\left(1-\frac{k}{n}-\frac{\alpha}{2\pi n}\right)$.
Assuming that this type of factorisation of the composite branch-point
twist field also holds in the off-critical theory we can obtain its
vacuum expectation value exploiting the results in Ref. \cite{DoyonBosonU1VEV}
\begin{equation}
\begin{split}\langle\mathcal{V}_{B}^{\varphi}\rangle= & \mathcal{N}\left(me^{\gamma_{E}}\right)^{\varphi(1-\varphi)/(2\pi)^{2}}\exp\left\{ -\frac{2}{\pi}\int_{0}^{\infty}\mathrm{d}t\frac{\sinh t\,\ln(\cosh t)}{\left(4t^{2}+\pi^{2}\right)\cosh^{2}t}\left[\vphantom{\frac{OOOOOO}{OOOOO}}\pi\cos\frac{\varphi}{2}\sinh\left(t\left(1-\frac{\varphi}{\pi}\right)\right)+\right.\right.\\
 & \qquad\qquad\qquad\qquad\qquad\qquad\qquad\left.\left.+2t\sin\frac{\varphi}{2}\cosh\left(t\left(1-\frac{\varphi}{\pi}\right)\right)\vphantom{\frac{OOOOOO}{OOOOO}}\right]\right\} ,
\end{split}
\end{equation}
with 
\begin{equation}
\mathcal{N}=\exp\left\{ -\frac{1}{3}(\gamma_{E}+\ln2)+\frac{1}{2}(1-\ln(2\pi))-\int_{0}^{\infty}\frac{\mathrm{d}t}{t}e^{-t}\left(\frac{\sinh t-t}{(e^{-t}-1)(\cosh t-1)}+\frac{1}{3}\right)\right\} \,,
\end{equation}
where $\gamma_{E}$ is Euler's constant.

Hence, for the VEV of the composite $U(1)$ branch-point twist field
in complex boson theory we have
\begin{equation}
\begin{split}\langle\mathcal{T}_{B,n}^{\alpha}\rangle=\mathcal{N}^{n} & \left(me^{\gamma_{E}}\right)^{\left(\frac{1}{6}\left(n-\frac{1}{n}\right)-\frac{\alpha^{2}}{(2\pi)^{2}n}+\frac{\left|\alpha\right|}{(2\pi)n}\right)}\exp\left\{ -\frac{1}{2\pi}\int_{0}^{\infty}\mathrm{d}t\frac{\sinh t\,\ln(\cosh t)}{\left(4t^{2}+\pi^{2}\right)\cosh^{2}t}\times\right.\\
 & \left.\left[(2t-i\pi)\csc\left(\frac{\pi-2it}{2n}\right)\cos\left(\frac{-2i\pi(n+1)t+2it\left|\alpha\right|-\pi\left|\alpha\right|+\pi^{2}}{2\pi n}\right)+\right.\right.\\
 & \left.\left.+(2t+i\pi)\csc\left(\frac{\pi+2it}{2n}\right)\cos\left(\frac{\pi(\pi+2i(n+1)t)-(\pi+2it)\left|\alpha\right|}{2\pi n}\right)+\right.\right.\\
 & \left.\left.+(2t-i\pi)\csc\left(\frac{\pi-2it}{2n}\right)\sin\left(\frac{(\pi-2it)(\pi(n-1)+\left|\alpha\right|)}{2\pi n}\right)+\right.\right.\\
 & \left.\left.+(2t+i\pi)\csc\left(\frac{\pi+2it}{2n}\right)\sin\left(\frac{(\pi+2it)(\pi(n-1)+\left|\alpha\right|)}{2\pi n}\right)\right]\right\} \,.
\end{split}
\end{equation}
For $n=1$, this formula again equals the vacuum expectation value
of the standard $U(1)$ twist field \cite{DoyonBosonU1VEV}.

\end{document}